\documentclass[english,reprint,amsmath,amssymb,aps,pra,superscriptaddress,floatfix]{revtex4-1}

\usepackage{dcolumn}
\usepackage{bm}

\usepackage{lipsum}

\usepackage{tabularx}
\usepackage{graphicx}
\usepackage{babel}
\usepackage{layouts}
\usepackage{epstopdf}
\usepackage{physics}
\usepackage{enumitem}
\usepackage{dcolumn}
\usepackage{bm}
\usepackage{amsmath}
\usepackage{comment}
\usepackage{bbold}

\usepackage[unicode=true,pdfusetitle,
 bookmarks=true,bookmarksnumbered=false,bookmarksopen=false,
 breaklinks=false,pdfborder={0 0 0},backref=false,colorlinks=true]
 {hyperref}
 
 \usepackage[normalem]{ulem}

\usepackage{color}

\makeatletter
\@ifundefined{textcolor}{}
{%
 \definecolor{BLACK}{gray}{0}
 \definecolor{WHITE}{gray}{1}
 \definecolor{RED}{rgb}{1,0,0}
 \definecolor{GREEN}{rgb}{0,1,0}
 \definecolor{BLUE}{rgb}{0,0,1}
 \definecolor{CYAN}{cmyk}{1,0,0,0}
 \definecolor{MAGENTA}{cmyk}{0,1,0,0}
 \definecolor{YELLOW}{cmyk}{0,0,1,0}
 }

\newcommand{\bit}{\begin{itemize}}
\newcommand{\eit}{\end{itemize}}

\newcommand{\bea}{\begin{eqnarray}}
\newcommand{\eea}{\end{eqnarray}}

\newcommand{\be}{\begin{equation}}
\newcommand{\ee}{\end{equation}}

\definecolor{dgreen}{rgb}{0.0, 0.5, 0.0}

\makeatletter

\begin{document}

\title{Momentum-space Gravity from the Quantum Geometry and Entropy of Bloch Electrons} 

\author{Tyler B. Smith}
\author{Lakshmi Pullasseri}
\author{Ajit Srivastava}

\email{ajit.srivastava@emory.edu }
\affiliation{
   Department of Physics, Emory University, Atlanta, GA 30322, USA.
   }

\begin{abstract}
{Quantum geometry is a key quantity that distinguishes electrons in a crystal from those in the vacuum. Its study continues to provide insights into quantum materials, uncovering new design principles for their discovery. However, unlike the Berry curvature, an intuitive understanding of the quantum metric is lacking. Here, we show that the quantum metric of Bloch electrons leads to a momentum-space gravity. In particular, by extending the semiclassical formulation of electron dynamics to second order, we find that the resulting velocity is modified by a geodesic term and becomes the momentum-space dual of the Lorentz force in curved space. We calculate this geodesic response for magic-angle twisted bilayer graphene and show that moir\'e systems with flat bands are ideal candidates to observe this effect. Extending this analogy with gravity further, we find that the momentum-space dual of the Einstein field equations remains sourceless for pure states while for mixed states it acquires a source term that depends on the von Neumann entropy, for small entropies. We compare this stress-energy equation with the weak-field limit of general relativity and conclude that the von Neumann entropy is the momentum-space dual of the gravitational potential. Consequently, the momentum-space geodesic equation for mixed states is modified by a term resembling an entropic force. Our results highlight connections between quantum geometry, momentum-space gravity and quantum information, prompting further exploration of this dual gravity in quantum materials.}
\end{abstract}

\maketitle

\section{Introduction}
The study of electrons in solids is one of the central themes in condensed matter physics. 
While an electronic band theory of crystals was developed soon after the formalization of quantum mechanics and is tremendously successful at explaining many electronic and optical properties of solids, over the past few decades it has become clear that the traditional approach involving only the energy dispersion of electrons is incomplete.
Our understanding of the quantum Hall effect has brought to the forefront quantum geometric quantities of Bloch states, such as Berry curvature, which are key ingredients of several topological phases of matter~\cite{thouless1982quantized,haldane1988model,kane2005quantum,xiao2010berry,qi2008topological}.  The study of such geometric quantities is particularly relevant since their gauge-invariant nature often makes them physically observable. Indeed, Berry curvature has been a revolutionary concept in explaining a range of physical phenomena such as anomalous Hall effects~\cite{chang1996berry,sundaram1999wave}, electronic polarization and orbital magnetization in solids~\cite{xiao2005berry,shi2007quantum,thonhauser2005orbital,ogata2015orbital}, and adiabatic charge pumping~\cite{niu1990towards} to name a few. Besides single-particle properties, it also affects Coulomb-correlated bound states such as excitons in crystals~~\cite{srivastava2015signatures,zhou2015berry} and could play an important role in our understanding of interacting many-electron systems.

In addition to Berry curvature, another fundamental geometric quantity is the quantum metric, which measures the infinitesimal distance between Bloch states defined over the momentum-space \cite{berry1984quantal, provost1980riemannian, anandan1990geometry}. Although, Berry curvature and the quantum metric have the same physical dimension and are often combined into a complex-valued quantity called the quantum geometric tensor with Berry curvature (quantum metric) as its imaginary (real) part, the role of the latter in determining the electronic properties of crystals is far less studied. Recently, its importance has been identified in a wide variety of phenomena including localization of Wannier functions \cite{marzari2012maximally, resta2011insulating,marzari1997maximally}, superconductivity and other phenomena in flat bands systems \cite{peotta2015superfluidity, torma2018quantum, julku2021excitations, rhim2020quantum, chaudhary2021shift,liang2017band,YuPing2021PRB,rossi2021quantum}, nonlinear response \cite{orenstein2021topology, shi2021geometric, morimoto2016topological, gao2014field, gao2019nonreciprocal, kozii2021intrinsic, ahn2021riemannian, julku2021excitations,sodemann2015quantum,topp2021light}, fractional Chern insulators and other topological phases~\cite{claassen2015position, haldane2011geometrical,Palumbo2018EPL,Palumbo2020PRX,lin2021band}, current noise \cite{neupert2013measuring}, magnetic susceptibility \cite{piechon2016geometric}, quantum phase transitions \cite{zanardi2008quantum, ma2010abelian,rezakhani2010intrinsic,zanardi2007information}, excitonic fine structure \cite{srivastava2015signatures, zhou2015berry}, and has been experimentally measured in photonic and atomic systems \cite{gianfrate2020measurement, ozawa2018extracting}.

Much progress has resulted through the identification of Berry curvature as a momentum-space dual of the magnetic field, and by exploiting this duality~\cite{price2014quantum, price2015four, fang2003anomalous,ozawa2015momentum}. However, such a fruitful analogy seems to be missing for the quantum metric. For example, the semiclassical equations of motion for electron dynamics in a band identifies the role of the Berry curvature, however, its cousin, the quantum metric is conspicuously missing. One can ask whether its status as a Riemannian metric leads to any analogies to Einstein's general relativity (GR), and if so, could it be investigated as momentum-space gravity. While such an analogy is worth investigating in its own right, it might also provide insights into real-space gravity and offer quantum condensed matter systems as testing ground for predictions of GR, starting from a fully quantum setting. Moreover, intuition gained from such an analogy can be exploited to engineer artificial gravity~\cite{HaldaneGraviton2012PRL,Golkar2016JHP,davis2021geodesic,kirmani2021realizing,wilson2020analogue}, in addition to artificial gauge fields, and affect the dynamics of quasiparticles.

Motivated by this question, in this work, we focus on two fundamental equations of GR, viz. the geodesic equation~\cite{carroll2019spacetime}
\begin{equation}
\hbar\dot{k}_\lambda = m \Gamma_{\mu \nu \lambda} \dot{x}^\mu \dot{x}^\nu,
\label{geodesicreal}
\end{equation}
and the Einstein field equations (EFE) 
\begin{equation}
R_{\mu \nu} - \frac{1}{2} R g_{\mu \nu} = T_{\mu \nu},
\label{EFEreal}
\end{equation}
and establish their momentum-space duals. Eq.~\ref{geodesicreal} describes the gravitational (pseudo-)force on a particle due to curved spacetime, as characterized by the real-space metric $g_{\mu \nu}$ and its Christoffel symbols $\Gamma_{\mu \nu \lambda} = \frac{1}{2}\left( \partial_\mu g_{\nu \lambda} +\partial_\nu g_{\lambda \mu} - \partial_\lambda g_{\mu \nu} \right)$. On the other hand, Eq.~\ref{EFEreal} relates the curvature of spacetime, captured by Ricci tensor ($R_{\mu \nu}$) and Ricci scalar ($R$), to the stress-energy tensor ($T_{\mu \nu}$), which acts as its source.
Eq.~\eqref{geodesicreal} describes the effect of curved spacetime on matter, while Eq.~\eqref{EFEreal} explains the effect of matter on the spacetime itself.  To quote Wheeler:  “spacetime tells matter how to move; matter tells spacetime how to curve.”

In this context, one can ask - (\emph{a}) whether the quantum metric of Bloch bands affects the motion of charge carriers by obeying the momentum-space dual of Eqs.~\ref{geodesicreal}, and (\emph{b}) in relation to the EFE, is there a quantity dual to the stress-energy tensor, i.e., which physical quantity tells the momentum-space how to curve? 

\begin{figure}
 \includegraphics[width=0.99\linewidth]{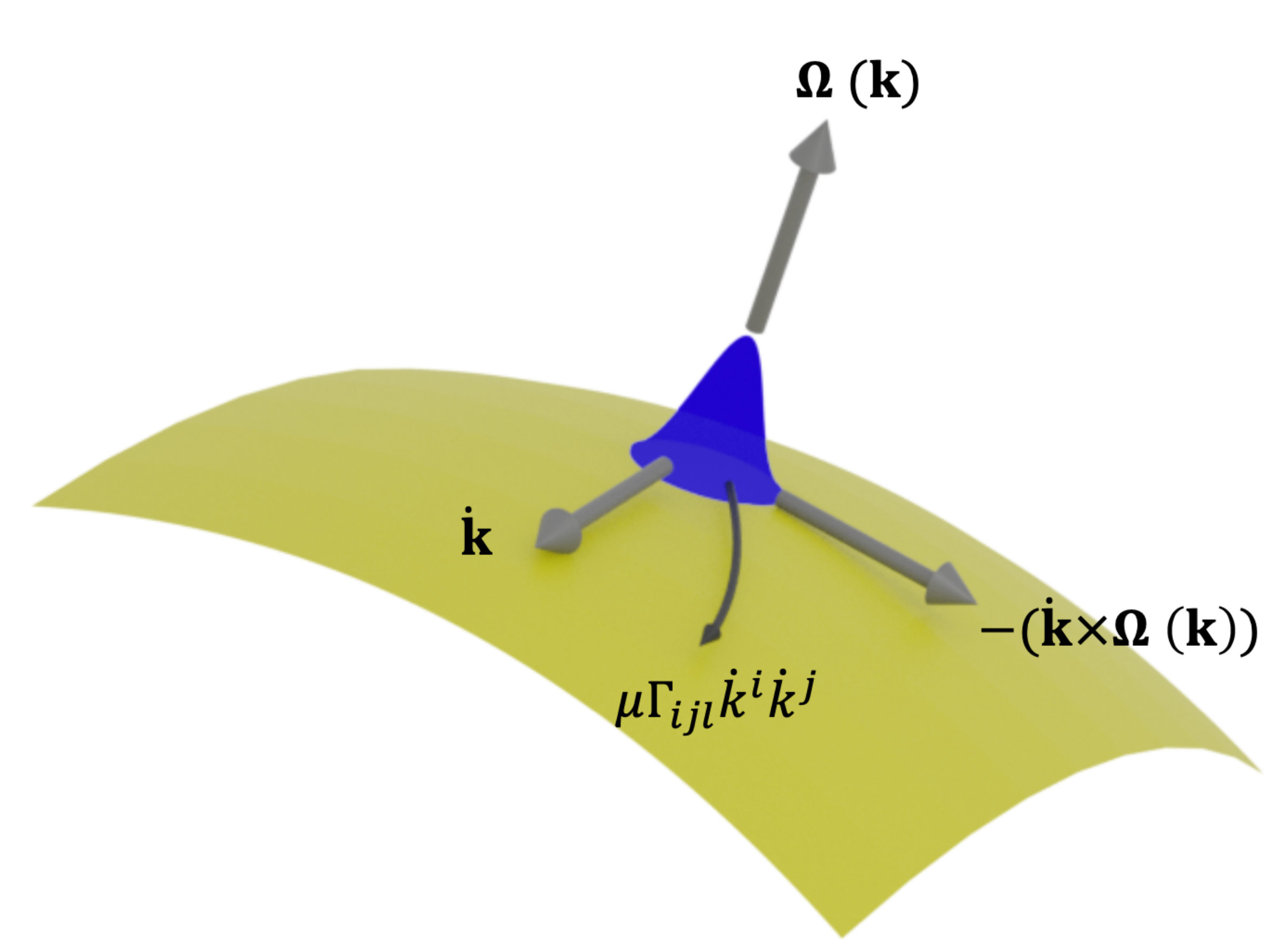}
  \caption{\textbf{ Semiclassical velocity is the momentum-space analog of the Lorentz force in curved spacetime.}
  The coupling between the electric field ($\dot{k}$) and the Berry curvature, $\Omega(\mathbf k)$, generates an anomalous velocity orthogonal to the applied field, while the quantum metric drives electrons in the direction of k-space geodesics via its Christoffel symbols, $\Gamma$ \eqref{kspacelorentz_final}. 
  The effects of $\Gamma$ are analogous to the effects of curved spacetime in general relativity, and the quantum metric can be seen as inducing momentum-space gravity.}
 \label{fig:fig1}
\end{figure}

Here, we address these questions by deriving the momentum-space dual of Eq.~\ref{geodesicreal} and \ref{EFEreal}. Our results can be summarized as follows: 

\begin{enumerate}
\item We find that the well-known semiclassical equation of motion for charge carriers in a Bloch band is modified in the presence of the quantum metric by a term which is the momentum-space dual of the geodesic equation. As shown in Fig.~\ref{fig:fig1}, under the semiclassical approximation, charge carriers respond not only to the Berry curvature but also to the quantum metric by moving along geodesics determined by the curvature of the momentum-space Bloch states. 

\item As for the EFE for the quantum metric, we first note that the quantum metric of pure states satisfies the vacuum EFE in arbitrary dimensions. 
Consequently, we consider mixed states over the bands with small von Neumann entropies. We find that in dimension $d$ = 2 EFE remain sourceless, while in $d \geq$ 3, a stress-energy tensor arises from the momentum-space Laplacian of the von Neumann entropy. This is analogous to the fact that the stress-energy tensor of spacetime can be expressed as the real-space Laplacian of a gravitational potential in the Newtonian limit of GR. 

In other words, for small entropy we recover a Newtonian limit of the EFE in the momentum-space, suggesting that, in our setting, entropy acts like a gravitational potential. As the von Neumann entropy is an entropy of entanglement, we conclude that severing of entanglement is responsible for the source term in the momentum-space EFE.

\item As a consequence of the finite von Neumann entropy in a mixed-state wave packet, we find that the momentum-space geodesic equation is modified by an additional term which resembles an entropic force.

\end{enumerate}

Our results seem to imply that the geometry of Bloch bands results in effective gravity from the quantum metric in addition to the effective gauge field (Berry curvature) and that this effective gravity affects the semiclassical dynamics of an electronic wave packet. The geodesic duality also clarifies why the quantum metric does not appear in traditional semiclassical wave packet dynamics in spite of having the same physical dimension as the Berry curvature. As discussed below, the geodesic duality should appear in the second-order response of charge carriers to an applied electric field, making its experimental realization feasible. Our predictions for the geodesic term are particularly relevant for charge carriers in quasi-flat bands where the group velocity term, of lowest order in applied electric field, becomes negligible. Such flat bands exist in moir\'e heterostructures of van der Waals materials such as magic-angle twisted bilayer graphene (MATBG) and transition metal dichalcogenides, making them an ideal platform to detect momentum-space geodesic dynamics~\cite{cao2018correlated,naik2018ultraflatbands,chaudhary2021shift}. 

The momentum-space EFE for the quantum metric, which are derived for a quantum condensed matter system, reveal intriguing connections between momentum-space gravity, quantum information, entanglement and GR. They appear similar in spirit to previous results connecting cosmological gravity with thermodynamics, especially entropy~~\cite{bekenstein1973black, hawking1975particle, ruppeiner1979thermodynamics, jacobson1995thermodynamics, padmanabhan2010thermodynamical, verlinde2011origin}. However, there are notable differences: (\emph{a}) Our derivation does not invoke entropy of black holes or the holographic principle, unlike previous works. (\emph{b}) Our results are valid for quantum systems at zero ``temperature" as the entropy involved is the von Neumann or the entanglement entropy and not necessarily the thermodynamic entropy. (\emph{c}) While our results are not concerning real gravity and its connection to entropy, they are easier to test in material systems and could offer valuable insights into the nature of the connection between gravity and quantum information~\cite{maldacena2013cool,van2010building}.

The remainder of this work is organized as follows. We start our analysis by first defining the quantum geometry arising from the Bloch bands in Sec.~\ref{sec:QGBloch}. This section has no new results but is included to make the discussion self-contained and to set the notation for the rest of the paper.  In Sec.~\ref{sec:geodesic}, we derive the second-order correction in electric field to the semiclassical equations of motion for a wave packet. The role of the quantum metric is identified by expressing the second-order corrections in terms of gauge-invariant, geometric quantities constructed from the Bloch states. We find that the momentum-space dual of the geodesic equation stems from the second-order correction to the energy of the wave packet. We use MATBG as an illustrative example to calculate the geodesic response and show that it is sizable for reasonable experimental parameters.

In Sec.~\ref{sec:EFE}, we consider the momentum-space dual of the EFE and find that the quantum metric for pure states satisfies the vacuum EFE. By generalizing the quantum metric of pure states to the Bures metric of mixed states, we find that, for small von Neumann entropies of the mixed state, the Bures metric can be thought of as a conformally scaled pure-state metric. The resulting momentum-space EFE then yield a source term which is proportional to the Laplacian of the von Neumann entropy. In Sec.~\ref{sec:Expt}, we briefly comment on possible ways to experimentally verify our predictions. Finally, in Sec.~\ref{sec:Conclusion}, we provide a summary of our work and identify future directions. Details of the analytical analysis are relegated to appendices.

\section{Quantum geometry of Bloch bands}\label{sec:QGBloch}
The quantum geometry of Bloch bands is defined in analogy with differential geometry by quantifying how the Bloch states $| u_n (\mathbf{k})\rangle$ vary smoothly with $\mathbf{k}$. The set of all inequivalent wave vectors $\mathbf{k}$ forms a closed manifold owing to the periodicity, $\mathbf{k} \equiv \mathbf{k} + \mathbf{G}$, where $\mathbf{G}$ is any reciprocal lattice vector. Therefore, in $d$-dimensions, such an identification makes the momentum space ${T}^d$, a $d$-dimensional torus over which the Bloch states are arranged. The most well-studied quantity that follows from this geometry is the Berry curvature for a band $n$, which is typically defined from a connection $\mathcal{A}_n(\mathbf{k}) = i \langle u_n(\mathbf{k}) | \mathbf{\nabla}_{\mathbf{k}} u_n(\mathbf{k}) \rangle$ as $\Omega_n(\mathbf{k}) = \nabla_{\mathbf{k}} \times \mathcal{A}_n(\mathbf{k})$. Berry curvature and its $\mathbf{k}$-space integral, Berry phase, are often interpreted as a momentum-space magnetic field and magnetic flux, respectively, while the Berry connection, $\mathcal{A}_n(\mathbf{k})$, acts as the corresponding vector potential. Another gauge-invariant quantity, which is of second order in k-space derivatives, is the infinitesimal distance between Bloch states $ds^2 = |\langle u_n(\mathbf k)|u_n(\mathbf k + d\mathbf k)\rangle|^2$. This distance can be expressed as $ds^2 = g_{ij}dk_i dk_j$, which defines $g_{ij}$ as the quantum metric. The components of the metric can be computed directly from the Bloch states, 
$ g_{ij}(\mathbf{k}) =  \mathrm{Re}[\langle \partial_{k^i}u_n(\mathbf k)|\partial_{k^j}u_n(\mathbf k)\rangle] -\mathcal{A}_{i}(\mathbf k) \mathcal{A}_{j}(\mathbf k) $.  
Remarkably, both $\Omega_n(\mathbf{k})$ and $g_{ij}(\mathbf{k})$ are second order in the momentum-space derivatives of $| u_n (\mathbf{k})\rangle$ and can be combined into real, symmetric and imaginary, antisymmetric parts of a larger Hermitian tensor known as the quantum geometric tensor (Fubini-Study metric), $\mathcal{Q}_{i j, n} = g_{ij,n} + i \Omega_{i j, n} /2$~\cite{provost1980riemannian}. In addition to a single band Berry connection, interband Berry connection is defined as $\mathcal{A}_{mn} = i \langle u_m(\mathbf{k}) | \mathbf{\nabla}_{\mathbf{k}} u_n(\mathbf{k}) \rangle$, where $m$ and $n$ are band indices with $m \neq n$.

\section{Momentum-space Geodesic equation}\label{sec:geodesic}
To explore how the quantum metric affects the carrier dynamics in solids, it is instructive to start with the semiclassical model of wave packet dynamics, which has been helpful in elucidating the role of Berry curvature in carrier dynamics~\cite{xiao2010berry}. In the presence of an external electromagnetic field, the center-of-mass momentum ($\mathbf{k}$) of a charged particle is given by the Lorentz force equation~\cite{misner1973gravitation},
\begin{equation}  
\dot{k}_\mu = \dot{x}^\nu \left( \frac{q}{\hbar} F_{\mu\nu} \right),
\end{equation}
where Greek indices run from 0, 1, \ldots, $d$,  $F_{\mu\nu}$ is the electromagnetic tensor and $x^\nu$ are the components of the position four-vector. It has been shown that the momentum-space dual of the above equation, obtained by making the analogy $x \leftrightarrow k$ and $\frac{q}{\hbar}F_{\mu\nu} \leftrightarrow \Omega_{\mu\nu}$, is
\begin{equation}
\dot{x}_\mu = \dot{k}^\nu \Omega_{\mu\nu}.
\end{equation}
The space (time) component of  $\Omega_{\mu\nu}$ is responsible for the anomalous Hall (group) velocity  stemming from the Berry curvature (energy dispersion).
We swap upper and lower indices in our analogy, as covariant quantities in spacetime are contravariant in momentum space.
This equation describes the semiclassical motion of the center-of-mass of a wave packet created from Bloch states of a given band $n$~\cite{chang2008berry}. As we are interested in the non-relativistic limit, in the following, we use Latin indices to label spatial coordinates. If we consider the more general setting of a curved space, the Lorentz force on a charged particle becomes:
\begin{equation}\label{realgeodesic}  
\dot{k}_a = \frac{q}{\hbar}E_a + \left( \dot{\mathbf{x}} \times \frac{q}{\hbar} \mathbf{B}\right)_a
+ \frac{m}{\hbar} \Gamma_{bca}\dot{x}^b\dot{x}^c,
\end{equation}
where the second term on the RHS is the ``gravitational force'' arising from the curvature of spacetime~\cite{carroll2019spacetime}. The first term on the RHS of Eq.~\ref{realgeodesic} can be expressed as $\frac{1}{\hbar}\partial_{x_a}\mathcal{E}(\mathbf{r})$. One is thus tempted to take the momentum-space dual of Eq.~\ref{realgeodesic} to describe the role of quantum metric in semiclassical wave packet dynamics,
\begin{equation}
\dot{x}_a \overset{?}{=} \frac{1}{\hbar}\partial_{k^a}\mathcal{E}(\mathbf{k}) + \left( \dot{\mathbf{k}} \times \mathbf{\Omega}\right)_a + \frac{\mu}{\hbar}\Gamma^{(k)}_{bca} \dot{k}^b\dot{k}^c,
\label{kspacelorentz}
\end{equation}
where $\Gamma^{(k)}_{bca}$ are the Christoffel symbols corresponding to quantum metric and $\mu$ is the momentum-space analog of mass. 

In order to arrive at the expression above, we consider the case of only an electric field such that $\dot{\mathbf{k}} = -\frac{e}{\hbar}\mathbf{E}$. The geodesic-like term then becomes second-order in the electric field, whereas the effect of Berry curvature is of first order. This suggests that one must extend the semiclassical formulation of electron dynamics to second-order in the external field in order to capture the role of the quantum metric. As there have been previous studies on the second-order response to electric and magnetic fields~\cite{gao2014field,kozii2021intrinsic, gao2019nonreciprocal, sodemann2015quantum,Xiao2021PRBthermo,Xiao2021PRBmag}, the scope of our study is to highlight the role of the quantum metric with regards to the momentum-space geodesic term of Eq.~\ref{kspacelorentz}.

\subsection{Dressed Hamiltonian to the second order in the velocity gauge}
In order to correct the semiclassical wave packet dynamics up to second order in $\mathbf{E}$, we restrict ourselves to a gapped, two band model with Hamiltonian $H(\mathbf{k})$ parameterized by the crystal momentum, $\mathbf{k} $. In the presence of an electromagnetic perturbation given by a potential $\mathbf{A}(\mathbf{x},t)$, the dressed Hamiltonian in the velocity gauge is $\tilde{H} = H(\mathbf{k} - \mathbf{A})$, where we have set $|e|$ = 1 and $\hbar$ = 1. Although the semiclassical equations consider a time-independent electric field, it is instructive to consider an AC electric field and then take the zero frequency limit to recover the DC result. Indeed, even for a DC electric field, a time-dependent treatment is necessary because a purely Hamiltonian dynamics does not attain a steady-state. Therefore, we assume that the steady-state is achieved through phenomenological scattering processes which are ubiquitous in a real crystal, and derive this steady-state, DC limit of the AC result. Furthermore, we assume a spatially homogeneous and harmonic in time electric field, $\mathbf{E}(t) = \mathbf{E}_0 \mathrm{cos} (\omega t)$ such that $\mathbf{A}(t) = -\mathbf{E}_0\frac{\mathrm{sin} (\omega t)}{\omega}$. 
Thus, the vector potential (in this gauge) corresponds to a gauge-invariant, observable quantity.

The bare Hamiltonian can be written as $ H = \sum \limits_{n} \langle u_n | H | u_n \rangle  | u_n \rangle \langle u_n | $, where the band index $n$ = 0, 1 and dependence on $\mathbf{k}$ is suppressed. The matrix elements of the dressed Hamiltonian in the unperturbed basis are 
\begin{equation} 
\langle u_m | \tilde{H} | u_n \rangle = \tilde{\mathcal{E}}_0 \langle u_m | \tilde{u}_0 \rangle \langle \tilde{u}_0 | u_n \rangle +  \tilde{\mathcal{E}}_1 \langle u_m | \tilde{u}_1 \rangle \langle \tilde{u}_1 | u_n \rangle.
\end{equation}
This expression is exact with $\tilde{\mathcal{E}}_{0,1}$ and $| \tilde{u}_{0,1} \rangle$ being the energies and eigenstates of $\tilde{H}$, respectively.

The elements of the dressed Hamiltonian in the unperturbed eigenbasis depend on the Bloch overlaps, which can be expanded to calculate $\tilde{H}$ up to second order in $\mathbf{A}$ (see Appendix \ref{Appendix B}). In order to express the perturbative corrections in terms of the gauge-invariant, geometric quantities of the Bloch states, we make use of the following relations(see Appendix \ref{Appendix A}):
\begin{eqnarray}
|\langle u_n | \tilde{u}_n \rangle |^2 &\approx& 1 - g_{ij}A^iA^j, \\
|\langle u_m | \tilde{u}_n \rangle |^2 &\approx& g_{ij}A^iA^j, \quad m \neq n,
\end{eqnarray}
where `$\approx$' stands for `equal up to second order'. As shown in the Appendix \ref{Appendix B}, the off-diagonal matrix elements of $\tilde{H}$ contain interband Berry connections $\mathcal{A}_{mn}$, which can be related to gauge invariant quantities by the following identity:
\begin{equation}
\sum_{n \neq 0} \mathcal{A}_{0n} \mathcal{A}_{n0} = g_{ij,0} - \frac{i}{2}\Omega_{ij,0},
\label{identity}
\end{equation}
where $n$ = 0 labels the band from which the electronic wave packet is made.

To isolate the effect of the Bloch geometry and focus on the role played by the quantum metric in the geodesic duality, we assume negligible dispersion of the bands by setting $\partial_{k_i} \mathcal{E}_n$ $\sim$ 0 for $n = 0,1$. This assumption is also motivated by current studies on MATBG and other moir\'e systems featuring flat bands wherein quantum geometry plays a dominant role~\cite{}.
We emphasize that while a two-band Hamiltonian can always be ``flattened" ($\partial_i \mathcal{E}_n = 0$) 
without affecting the Bloch band geometry, we are not restricting ourselves to such a case. Indeed, the role of dispersion can be easily considered in our treatment, but as it is not the main focus of this paper, we choose not to emphasize it.

\subsection{Second order corrections to the semiclassical dynamics of electronic wave packet}
Using the dressed Hamiltonian, we can apply the standard techniques of time-dependent perturbation theory to correct the wave function and its associated energy. This in turn gives us the corrections to Berry connection and Berry curvature up to second-order in the electric field. Let $\mathbf{a}'$ and $\mathbf{a}''$ be the first and second corrections to the Berry connection $\mathcal{A}_0 (\mathbf{k})$, respectively. The gauge-invariant quantities that could enter the semiclassical equation for the wave packet dynamics are their 
derivatives. As $\mathbf{\Omega} '' = \nabla_{\mathbf{k}} \times \mathbf{a}''$ is the second-order correction to the Berry curvature, which enters the semiclassical equation of motion as $\mathbf{\Omega} \times \mathbf{E}$, its contribution is third order in the electric field and hence is dropped in our analysis. 
To find $\mathbf{\Omega} '$ and its associated response, we first calculate $\mathbf{a}'$ (see Appendix \ref{Appendix B}):

\begin{equation}
a'_{i}(\mathbf k) =  \frac{\Delta}{ \omega (\Delta^2 - \omega^2 )}E^j_0 \left(  g_{i j}  S^{\omega} (t) + \frac{\Omega_{ij}}{2} A^{\omega} (t) \right),
\end{equation}
where $\Delta$ is the band gap, $S^\omega $ ($A^\omega $) is a symmetric (antisymmetric), $\mathbf{k}$-independent factor that is harmonic in $\omega$. Thus, we find that the first-order correction to the Berry connection is gauge invariant, involving QGT and Berry curvature. The corrected Berry curvature then becomes $\tilde{\Omega} = \Omega + \nabla_{\mathbf{k}} \times \mathbf{a}'$.

Similarly, we arrive at the expression for $\mathbf{a}''$ which is presented in Appendix . Unlike $\mathbf{a}'$, $\mathbf{a}''$ involves the interband Berry connection, $\mathcal{A}_{01}$, as well. In addition to the correction to the Berry curvature, the AC results give the first (second) order correction to group velocity as $\partial_t \mathbf{a}'$ ($\partial_t \mathbf{a}''$). As expected, the first-order corrections result in a response at $\omega$ whereas the second-order correction results in a nonlinear response at DC and 2$\omega$.
Finally, the energy of the wave packet to second order is given by -
\be
\tilde{\mathcal{E}} = \mathcal{E}_0 + \Delta \beta (t) g_{ij} E_0^i E_0^j.
\ee
We note that the energy correction to the wave packet is second order in $\mathbf{E}_0$ and is proportional to the quantum metric. The $\mathbf{k}$-space gradient of the above term gives an additional correction to the ``group velocity" part of the semiclassical equations. 

With the above corrections to the Bloch band geometry, we are ready to write down the semiclassical equations of motion to second order in $\mathbf{E}_0$ as follows:
\be
\dot{\mathbf{k}} = - \mathbf{E},
\ee
\be
\dot{\mathbf{r}} = \nabla_{\mathbf{k}} \tilde{\mathcal{E}} (\mathbf{k})+\dot{\mathbf{k}} \times \tilde{\mathbf{\Omega}} (\mathbf{k}) -\partial_t (\mathbf{a}' + \mathbf{a}''),
\label{semiclass1}
\ee
where we have taken the sign of the charge carrier to be negative in order to describe electrons.

\subsection{Second order semiclassical equations in the steady-state and DC limit}
Next, we consider the steady-state limit together with the DC limit, $\omega \to 0$. As mentioned earlier, in absence of dissipation from impurity scatterings, there is no steady-state solution even in the case of DC electric field. The role of impurity scattering then is to average the rapidly oscillating terms at e$^{i\Delta t}$ to zero. In other words, for timescales larger than the characteristic scattering time $\tau$, one can drop the harmonic terms in $\Delta t$ and then take the limit $\omega \to 0$ to obtain the DC results. As shown in the Appendix \ref{Appendix B}, for the first-order correction to Berry connection, we obtain
\be
a'_i (\mathbf{k}) \to -\frac{2}{\Delta}g_{ij}E_0^j,
\label{a'}
\ee
which coincides with the first-order positional shift derived by Gao \emph{et al}. using time-independent perturbation theory~\cite{gao2014field,gao2015geometrical}. Furthermore, as shown in the Appendix \ref{Appendix A}, $\partial_t \mathbf{a}'\to 0$ and $\partial_t \mathbf{a}'' \to 0$, leaving only the $\tilde{\mathbf{\Omega}}$ and $\nabla_{\mathbf{k}} \tilde{\mathcal{E}}(\mathbf{k})$ terms.
The time-dependent factor $\beta (t)$ appearing in $ \tilde{\mathcal{E}}(\mathbf{k})$ reaches a steady-state value of
\be
\beta \to \frac{2}{\Delta^2}.
\ee

Thus, the correction to group velocity term in the steady-state, DC limit becomes -
\begin{eqnarray}
    \nabla_\mathbf{k} \tilde{\mathcal{E}}_l &=& \frac{2}{\Delta}\left(\partial_{k^l} g_{ij}\right) E_0^i E_0^j \\
     &=& \mu \Gamma_{ijl} E_0^i E_0^j,
\end{eqnarray}
where $\mu = 4/\Delta$, $\Gamma_{ijl}$ = $\frac{1}{2} \left(\partial_{k^j} g_{li} + \partial_{k^l} g_{ij} - \partial_{k^i} g_{jl} \right)$ and we have used the symmetry of $E_0^i E_0^j$ under $i \leftrightarrow j$.

With this, we finally arrive at the semiclassical equation for wave packet dynamics to second order in $\mathbf{E}_0$ under the steady-state, DC limit:

\be
\dot{r}_l = (\dot{\mathbf{k}} \times \tilde{\mathbf{\Omega}})_l + \mu
\Gamma_{ijl} \dot{k}^i \dot{k}^j.
\label{kspacelorentz_final}
\ee
This should be compared to the actual Lorentz force equation for an electron in curved space:
\be
\dot{k}_l = -\frac{e}{\hbar}(\dot{\mathbf{r}} \times \mathbf{B})_l + m \Gamma_{ijl}\dot{r}^i \dot{r}^j.
\ee
Thus, we find that the second order semiclassical equation becomes the $\mathbf{k}$-space dual of the Lorentz force equation in curved space, with the momentum-space quantum metric playing the role of the classical spacetime metric. The real-space mass becomes equivalent to $\mu = 4/ \Delta$. Thus, it is the second-order correction to the wave packet energy which becomes the geodesic term.

We also note that when $\omega \neq 0$ but $\omega \ll \Delta$, the $\partial_t(\mathbf{a}' + \mathbf{a}'')$ term will lead to linear and nonlinear responses at DC, $\omega$ and $2\omega$.  

\subsection{Steady-state analysis in the length gauge}
In this section, we independently check our steady-state, DC limit by analyzing the response to a constant electric field using the length gauge. We begin with a Hamiltonian parametrized by crystal momentum and perturb it with a constant, position independent electric field which couples to electrons via a dipole term:
\be
\tilde{H}(\mathbf{k}) = H(\mathbf{k}) + \mathbf{E} \cdot \mathbf{r}, 
\ee
where the position operator, $\mathbf{r}$, can be represented in the Bloch wave basis as~\cite{sundaram1999wave, xiao2010berry}
\be
\langle u_m(\mathbf k ) |\mathbf{r} | u_n(\mathbf k) \rangle = \mathcal{A}_{mn}(\mathbf{k}) \ \   \text{ if }  m \neq n.
\ee
Using this representation of the position operator, the perturbative correction to our two-band Hamiltonian can be represented as
\be
 H'  \approx  {\begin{pmatrix} \mathbf{E} \cdot\langle u_0(\mathbf k ) |\mathbf{r} | u_0(\mathbf k) \rangle  &  \mathbf{E} \cdot \mathcal{A}_{01}(\mathbf k)  \\\   \mathbf{E} \cdot \mathcal{A}_{10}(\mathbf k)  & \mathbf{E} \cdot \langle u_1(\mathbf k ) |\mathbf{r} | u_1(\mathbf k) \rangle  \end{pmatrix}}.
\ee
In this gauge, we can use time-dependent perturbation theory to compute the first-order correction to the wave function and Berry connection (see Appendix \ref{Appendix B}).  For the Berry connection, we find 
\be
{a}'_i = \frac{-2 g_{ij} E^{j}}{\Delta},
\ee
which coincides with Eq.~\ref{a'}. 
The second-order correction to the energy of the wave packet becomes (see Appendix \ref{Appendix B}):
\be
\mathcal{E}'' = \frac{2g_{ij}E^iE^j}{\Delta},
\ee
which yields a value of $\mu$ identical to that obtained in the previous section. We note that the second-order correction to the energy obtained above is consistent with the change in dipole energy from $\mathbf{E}\cdot \mathbf{r}$ to $\mathbf{E}\cdot (\mathbf{r} - \mathbf{a}')$ due to the positional shift  $\mathbf{a}'$.

\subsection{An illustrative model of MATBG for the geodesic equation}

\begin{figure}
 \includegraphics[width=0.999\linewidth]{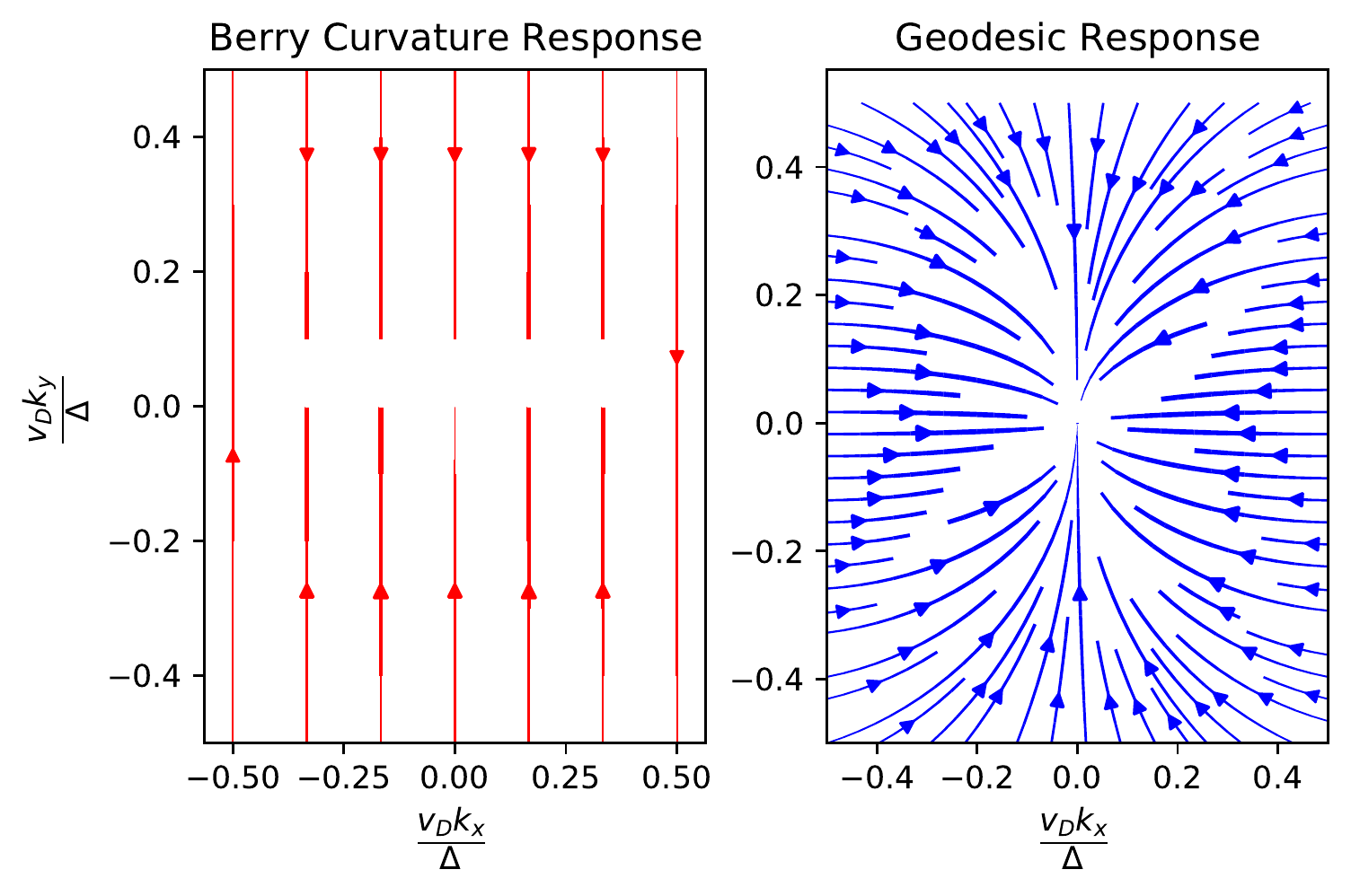}

  \caption{\textbf{ Berry curvature and the quantum metric produce distinct nonlinear responses. }
  While corrections to the Berry curvature produces a nonlinear Hall response, the ``geodesic response'' from the metric connection can be both parallel and orthogonal to the applied field.  Plots show second-order responses to an electric field in the $x$ direction as a function of dimensionless momentum. Line widths in the stream plots are proportional to the local magnitude of the response. }
 \label{fig:fig2}
\end{figure}  

\begin{figure}
 \includegraphics[width=0.999\linewidth]{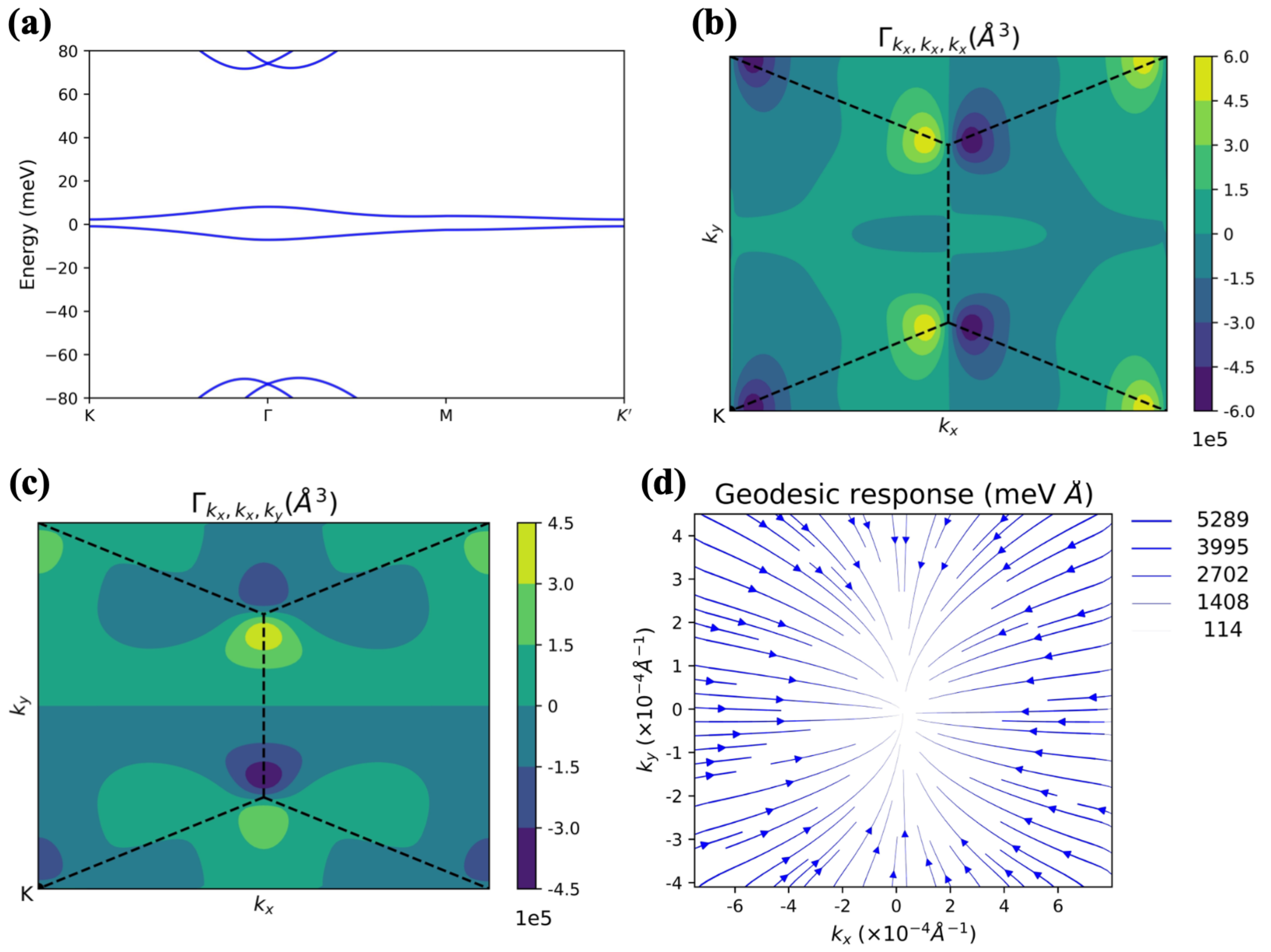}

  \caption{\textbf{ Geodesic response for MATBG }
\textbf{(a)} Energy dispersion obtained from the Hamiltonian in Eq.~\ref{MATBGH} describing MATBG, with a twist angle $\theta$ = 1.05$^\circ$ and $\Delta$ = 2 meV, showing two isolated quasi-flat bands. \textbf{(b)-(c)} Momentum-space Christoffel symbols calculated from the quantum metric and plotted in the moir\'e Brillouin zone (dashed lines mark the boundary of Brillouin zone). \textbf{(d)} Geodesic response or the second order velocity calculated near the $K$-point for an applied electric field of 1 V/$\mu$m along the $x$ direction. The range of momenta is chosen such that the deviation from two-band approximation is less than 15 $\%$.}
 \label{fig:fig2}
\end{figure} 
We begin by considering a two-band massive Dirac model of gapped graphene, which we ``flatten" to make the energy dispersion vanish. This toy-model provides qualitative insight into the behavior of the geodesic response. Subsequently, we use a more realistic model of MATBG with quasi-flat bands to calculate the geodesic response. 

Consider a modified two-band model of gapped graphene with the following Hamiltonian~\cite{srivastava2015signatures}:

\begin{equation} 
H(\mathbf {k})  = \frac{\Delta}{2\mathcal{E}(\mathbf{k})} {\begin{pmatrix}\Delta/2&v_D(\tau k_{x}-i{k_y})\\v_D(\tau k_{x}+i{k_y})&-\Delta/2 \end{pmatrix}}, \end{equation}
where $\tau = \pm 1$ is the valley index, $\Delta$ is the band gap, $v_D$ is the Dirac velocity and $\mathcal{E}(\mathbf k) = \pm (v_D^2|\mathbf{k}|^2 + (\Delta/2)^2 )^{1/2}$. The energy of each band is $\pm \Delta/2$, leading to non-dispersing bands. We note that this ``flattening" of the energy dispersion does not affect the quantum geometry of the bands which remains the same as that for the gapped graphene model.

The Berry curvature in each band has only a $z$-component (pointing out of the plane) :

\begin{equation} 
\Omega(\mathbf {k})  = \mp \frac{v_D^2\Delta}{4 (v_D^2|\mathbf{k}|^2 + (\Delta/2)^2)^{3/2}},
\end{equation}

and the components of the metric are found to be:
\begin{equation}  g_{i j}(\mathbf {k}) =  \frac{v_D^2\left( {(v_D^2|\mathbf{k}|^2 + (\Delta/2)^2)\delta_{ij} -  v_D^2 k_i}{k_j}  \right)}{4 (v_D^2|\mathbf{k}|^2 + (\Delta/2)^2 )^2}. \end{equation}

We can use the metric to directly compute the Christoffel symbols of the first kind.  
Assuming the applied field is in the $x$-direction, the response to the field is determined by  $\Gamma_{k_x k_x k_x} =  \frac{1}{2} \partial_{k_x} g_{x x}$ and $\Gamma_{k_x k_x k_y} =  \frac{1}{2} \partial_{k_y} g_{x x}$, which can be expressed as:
\begin{equation}
\begin{aligned}
\Gamma_{k_x k_x k_x} =    \frac{-v_D^4k_x( v_D^2k_y^2 + (\Delta/2)^2 )}{2(v_D^2|\mathbf{k}|^2 + (\Delta/2)^2 )^3}, \\
\\
\Gamma_{k_x k_x k_y} = \frac{- v_D^4k_y (v_D^2k_y^2 - v_D^2k_x^2 + (\Delta/2)^2)}{4(v_D^2|\mathbf{k}|^2 + (\Delta/2)^2 )^3}.
\end{aligned}
\end{equation}
$\Gamma_{k_x k_x k_x}$ determines the parallel component of the second-order geodesic response, while $\Gamma_{k_x k_x k_y}$ controls the geodesic response orthogonal to the applied field (nonlinear Hall-type response). Fig.~\ref{fig:fig2} shows the second-order contributions to the velocity.

For $v_D^2|\mathbf{k}|^2 \ll \Delta^2$, the strength of the geodesic response increases linearly with the magnitude of the momentum vector $\mathbf{k}$.
This can be contrasted with the response due to Berry curvature, which is effectively constant in this regime.
The response due to the dispersion will also depend linearly on $\mathbf{k}$ in this limit, but will be independent of the field strength.
The change in the anomalous velocity with $\mathbf{k}$ and $\mathbf{E}$ should thus reveal the role of the QGT connection in the second-order response of electrons.

Next, we consider MATBG which has negligible dispersion of bands and serves as an ideal system to highlight the role of quantum geometry. We use the tight-binding model of Koshino \emph{et al}., with a small sublattice gap $\Delta$ which serves as the band gap~\cite{Koshino2018PRX}. The single-particle Hamiltonian for TBG in the basis of ($A_1$,$B_1$,$A_2$,$B_2$) for the +$K$ valley reads:

\begin{equation}
   \mathbf{\mathcal{H}} = 
\begin{pmatrix}
    H_{1} & U^{\dagger} \\
    U & H_2 \\
\end{pmatrix} 
\label{MATBGH}
\end{equation}
where $H_{1/2} = -\hbar v [R(\pm \frac{\theta}{2})](\boldsymbol{k} - \boldsymbol{K^{1/2}}) \cdot (\sigma_x , \sigma_y) + \Delta \sigma_z$ and

\begin{equation*}
    U = U_1 + U_2 + U_3
\end{equation*}
with,
\begin{equation}
    U_1 = 
\begin{pmatrix}
    u & u' \\
    u' & u \\
\end{pmatrix} 
\end{equation}

\begin{equation}
U_2 =
\begin{pmatrix}
    u & u' e^{-i 2 \pi/3} \\
    u' e^{i 2 \pi/3} & u \\
\end{pmatrix} e^{i \boldsymbol{G}_1^M \cdot \boldsymbol{r}} 
\end{equation}

\begin{equation}
U_3 = 
\begin{pmatrix}
    u & u' e^{i 2 \pi/3} \\
    u' e^{-i 2 \pi/3} & u \\
\end{pmatrix} e^{i (\boldsymbol{G}_1^M + \boldsymbol{G}_2^M) \cdot \boldsymbol{r}}
\end{equation}
    

The momentum $\boldsymbol{k}$ in the given Hamiltonian is in the BZ of the original graphene layers and, $\boldsymbol{K^{1}}$ and $\boldsymbol{K^{2}}$ are the points where the original layers' Dirac points are located. Here, $\boldsymbol{G}_1^M$ and $\boldsymbol{G}_2^M$ are the reciprocal lattice vectors of the moir\'e BZ, which can be expressed in terms of the monolayer reciprocal lattice vectors $\{ \boldsymbol{G}_i \}$ as, $\boldsymbol{G}_i^M = R(-\frac{\theta}{2})\boldsymbol{G}_i - R(\frac{\theta}{2})\boldsymbol{G}_i$. 

We consider $\theta$ = 1.05$^\circ$ and $\Delta$ = 2 meV, which yields quasi-flat energy dispersion as shown in Fig.~3a. The modified Dirac velocity, $\tilde{v}_D$ $\sim$ $v_D$/2667, quantifies the flatness of bands in this moir\'e system. The lowest energy conduction and valence bands serve as isolated two bands of our analysis. Next, we calculate $\Gamma_{xxx}$ and $\Gamma_{xxy}$ from the quantum metric which are plotted in Fig.~3b-c. We verify the validity of two-band approximation using Eq.~\ref{identity} by restricting the summation to these bands and find a deviation $<$ 15$\%$ near $K$-point. The geodesic response in this range of momentum is shown in Fig.~3d for an applied electric field of 1 V/$\mu$m. As expected, the geodesic response of MATBG has qualitatively similar behavior compared to the toy-model in Fig.~2. We discuss possible experimental verification of the geodesic term in Sec.~\ref{sec:Expt}.

\section{ Momentum-space Einstein Field Equations}{\label{sec:EFE}}
\subsection{Pure states and vacuum EFE}

When quantum states are parameterized by crystal momentum, the momentum space inherits the metric of the underlying Hilbert space.  
The space of all pure states in Hilbert space can be seen as a high dimensional sphere, while the space of physically distinguishable states is the quotient space obtained by identifying all states on the sphere that differ by a phase factor.
This space of physically distinct quantum states is known as the projective Hilbert space of states, and has the geometry of a complex projective space.
All such complex projective spaces possess a canonical Riemannian metric, known as the Fubini-Study metric~\cite{moroianu2007lectures}.  
This Fubini-Study metric is Einstein, i.e., it has a Ricci tensor proportional to itself.
As a result, the quantum metric, which is a Fubini-Study metric, is a vacuum solution of the Einstein field equations in the projective Hilbert space~\cite{besse2007einstein}: 
\begin{equation} R_{ij} - \tfrac{1}{2}R \, g_{ij} + \Lambda g_{ij} = 0,
\label{EFE2}
\end{equation}
where $R_{ij}$ is the Ricci tensor, $R$ is the scalar curvature, and $\Lambda$ is the cosmological constant.

The momentum space can be thought of as being embedded in the Hilbert space: it parameterizes a submanifold of quantum states $| u_0(\mathbf k)\rangle$.
It also inherits the (pullback) metric of the Hilbert space, defined via the overlaps of $k$-dependent Bloch functions: $g_{ij}dk^i dk^j = 1 - |\langle u_0(\mathbf k)|u_0(\mathbf k + \mathbf{dk}) \rangle |^2$ ~\cite{footnote}.

\subsection{Mixed states, Bures metric and the source of EFE}
One can generalize the notion of the quantum metric for pure states to mixed states where it is called the Bures metric~\cite{bures1969extension,dittmann1993riemannian}. Here we show that the Bures metric for mixed states can have a nonzero stress-energy tensor even when the pure state metric is a vacuum solution of the EFE.  
 Since all two-dimensional metrics have vanishing stress-energy by construction, we show this for an $N >2$ band quantum system in which the Bloch states have three dimensions of crystal momentum (see Appendix \ref{Appendix C}).  
 
First, we consider the density matrix of a mixed quantum state: 
\begin{equation}\rho(\mathbf{k})=  
\sum_{n=0}^{N} p_n(\mathbf k ) | u_n(\mathbf k) \rangle  \langle u_n(\mathbf k ) |, \label{densitymatrix}\end{equation}
where $ p_n (\mathbf k )$  is the probability for the system to be in $|u_n(\mathbf{k})\rangle$. The difference in the density matrices of the perturbed and unperturbed system can be expressed in terms of the Bures metric (see Appendix \ref{Appendix C}).

\begin{figure}
 \includegraphics[width=0.59\linewidth]{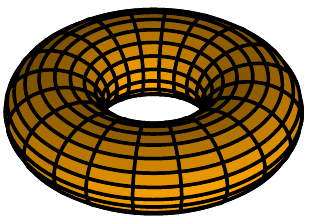}
 \includegraphics[width=0.59\linewidth]{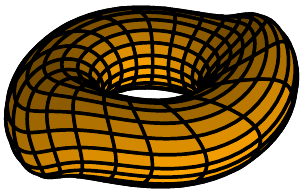}
  \caption{\textbf{ Entropy and entanglement deform the quantum geometry of the Brillouin zone.}
  Top: for pure states with zero entropy, the quantum distance between states with different momenta is defined via Bloch overlaps.  Bottom: for mixed states, the quantum metric \eqref{buresmain} is deformed by a conformal scale factor due to nonzero entropy.  This conformal factor preserves angles but alters the distance between points in momentum space.}
 \label{fig:fig3}
\end{figure}

Next, we assume - (a) the probabilities $p_n(\mathbf k ) $ change slowly in k-space (as shown in Appendix \ref{Appendix C}, it suffices if $\mathbf{\nabla}_k (\mathrm{ln} (p_n(\mathbf{k})) \sim 0)$, and (b) all the probabilities above the ground state are small, i.e., $p_0(\mathbf{k}) \gg p_i(\mathbf{k})$ for $i \neq 0$. Under these assumptions, we find that the Bures metric $\bar{g}_{i j}$ takes the following form (see Appendix \ref{Appendix C}):
\begin{equation} \bar{g}_{i j}dk^i dk^j  = e^{-\frac{S(\mathbf k)}{k_B} } g_{ i j}dk^i dk^j.  
\label{buresmain}
\end{equation}
We see that the Bures metric differs from the Fubini-Study metric by a conformal scale factor due to von Neumann entropy of the mixed state. 
The above equation \eqref{buresmain} is completely general, in that it holds for any number of parameters and any number of bands. This fact that the entropy of mixed states deforms the quantum geometry conformally is depicted as a  cartoon in Fig.~\ref{fig:fig3}.

The correction to Eq.~\ref{EFE2} arising from the conformal scaling of the quantum metric can be written as a source term of the momentum-space EFE (see Appendix \ref{Appendix C}). The trace of this momentum-space stress-energy tensor for the Bures metric is found to be:
\begin{equation} T =    -\frac{R}{2 k_B}S(\mathbf k) - \frac{1}{k_B} \Delta_{\mathbf k} S(\mathbf k)  .   
\label{stressenergy}
\end{equation}
As $R$ is a constant for a Fubini-Study metric, $S(\mathbf{k})$ being constant implies that $T$ is also a constant and can be absorbed into the cosmological constant. As a result, a non-constant entropic field is responsible for a non-trivial source term.
From the above equation, we can conclude that within the assumptions of small and slowly varying entropy, the momentum-space EFE acquires a source term due to the entropic field. The stress-energy in \eqref{stressenergy} is analogous to the weak field limit of general relativity in which Newtonian gravity and Poisson's equation are recovered~\cite{misner1973gravitation}. In other words, the von Neumann entropy is analogous to the gravitational potential of the Newtonian limit. The Bures metric responds to changes in entropy, just as the spacetime metric responds to changes in the distribution of matter. 

We note that the von Neumann entropy of the mixed state appearing above must arise as a result of entanglement of the electron with a larger system, such as a measurement apparatus or a thermal reservoir, whose dynamics are not of interest and are to be traced out. Thus, we can think of the source term in the momentum-space EFE to arise when such an entanglement is severed by averaging over environmental degrees of freedom. The entropy resulting from our lack of information in the larger system appears as a gravitational potential in the momentum-space.

\subsection{ Illustrative Example: Momentum-space EFE for 3D Dirac Fermion}
Assuming the charge carrier behaves as a three-dimensional Dirac fermion, we can use the momentum-space Dirac equation as its effective Hamiltonian:
\begin{equation}
H = 
  \begin{bmatrix}
    \Delta \mathbf{I} & \vec{\sigma}\cdot\mathbf{k}  \\
    \vec{\sigma}\cdot\mathbf{k} & -\Delta \mathbf{I}
  \end{bmatrix}.
\end{equation}
The resulting Bloch function is then described by a four-spinor:
\begin{equation}
|u_0(\mathbf k) \rangle =\textstyle\frac{|\mathcal{E}(\mathbf k)|+\Delta}{\sqrt{\mathbf k^2 + \left(|\mathcal{E}(\mathbf k)| +\Delta \right)^2}} \begin{bmatrix}
    \frac{\vec{\sigma} \cdot \mathbf k}{|\mathcal{E}(\mathbf k)| + \Delta} \chi_s \\ \chi_s 
\end{bmatrix},
\end{equation}
where the state has been normalized to one and $|\mathcal{E}(\mathbf k)|$ is the absolute value of the energy of the band, $\mathcal{E}(\mathbf k) = -\sqrt{\Delta^2 + \mathbf{k}^2}$.
While we deal with the negative energy solution here, it is straightforward to show that our expression for the metric also applies to the positive energy band.
Note that $\chi_s$ is an arbitrary two spinor, and the ground state of the Dirac Hamiltonian is thus degenerate. 

As shown by \citet{matsuura2010momentum}, we find the following expression for the quantum metric when $\mathbf k^2 \ll \Delta^2$ (see Appendix \ref{Appendix D}):
\begin{equation}
g_{ij} dk^i dk^j \approx 
 \frac{\delta_{ij} dk^i dk^j }{ 4\Delta^2}  \ \ \ \ \ \ \text{for } \mathbf k^2 \ll \Delta^2. 
\label{dirac_metric}
\end{equation}

In this limit, the pure state metric becomes completely flat and is thus a vacuum solution of the EFE.
For a mixed state with finite entropy, the stress-energy in then described by \eqref{stressenergy} with $R=0$:
\begin{equation} T \approx    - \frac{1}{k_B} \Delta_{\mathbf k} S(\mathbf k)  .   
\label{stressenergy_dirac}
\end{equation}
We see that the analogy between the weak field limit of GR and our momentum-space expressions for small entropy becomes even more striking in this regime, with the entropy and gravitational potential acting as a source of stress-energy in momentum space and spacetime respectively.

\subsection{Entropy maximization}
Next, we study how the geodesic equation is modified for mixed states. We can understand the conformal scaling in \eqref{buresmain} by viewing the resulting change in the momentum-space geodesic term. As shown in the Appendix~\ref{Appendix C}:
\begin{equation} 
\Gamma_{ijl}E_0^i E_0^j \rightarrow  \bar{\Gamma}_{ijl}E_0^i E_0^j -\frac{1}{2k_B} (\partial_{k^l}S(\mathbf{k})) \bar{g}_{ij}  E_0^i E_0^j,
\label{connection}
\end{equation}
where $\bar{\Gamma}_{ijl} = e^{-S(\mathbf{k})/k_B} \Gamma_{ijl}$.
Thus, in addition to the conformal scaling of the original geodesic equation, additional terms proportional to gradient of the entropy appear, which 
can be thought of as a momentum-space entropic force. We can thus understand the paths of least distance for mixed states as having two components:
one attempting to minimize the distance associated with the underlying pure states, and another trying to maximize the entropy of the mixed state.

\section{Experimental considerations}{\label{sec:Expt}}
\subsection{Geodesic term}
Our analysis shows that within the two-band approximation, the semiclassical equations of motion for Bloch electrons feature a geodesic term which is of second-order in the electric field. Calculating the geodesic response for a realistic material requires the inclusion of scattering processes and is beyond the scope of this work. In the following, we comment on the size of the geodesic term and possible materials systems where our predictions can be tested. 

The ``non-flattened" part of the toy-model of Sec. \ref{sec:geodesic}, is applicable to bilayer graphene in an out-of-plane electric field (which opens up a band gap), or semiconducting transition metal dichalcogenides (TMDs). We first estimate the size of the second-order geodesic term in comparison to the zeroth order group velocity term. Assuming an applied electric field in the $x$-direction, the group velocity is in the longitudinal direction, which we compare to the longitudinal part of the second-order velocity from the geodesic term. The dimensionless ratio of the two quantities can be expressed as -
\begin{equation}
    \frac{\dot{r}_x^{(0)}}{\dot{r}_x^{(2)}} = \frac{2v_D^2}{(\Delta/2)^4}\left(\frac{e}{\hbar}E_x\right)^2 \mathrm{cos}^3 \theta \left(\mathrm{sin}^2\theta \mathrm{cos}^2 \phi - 1 \right) ,
\end{equation}
where $\mathrm{cos}\theta = \Delta/2\mathcal{E}(\mathbf{k})$ and  tan$\phi = k_y / k_x$ correspond to angles on the Bloch sphere. For $v_D|\mathbf{k}| \ll \Delta/2$, $\theta \approx$ 0 and the ratio becomes -
\begin{equation}
    \frac{\dot{r}_x^{(0)}}{\dot{r}_x^{(2)}} = - \frac{2v_D^2}{(\Delta/2)^4}\left(\frac{e}{\hbar}E_x\right)^2.
\end{equation}
Taking $v_D \sim 10^6$ m/s and $\Delta \sim$ 10 meV, which are typical values for bilayer graphene in an out-of-plane electric field, and $E_x \sim$ 1 mV/$\mu$m, the two terms become comparable. Thus, with a reasonable electric field, the magnitude of the geodesic term can be made comparable to the group velocity term. As shown in Fig. \ref{fig:fig2}, the geodesic term makes the velocity point towards the $K$-points as opposed to the group velocity term, which can be probed in experiments. Here, we have neglected the second-order response from the finite energy dispersion which might compete with the geodesic term. However, when $v_D|\mathbf{k}| \gg \Delta/2$, the dispersion becomes almost linear and its second-order response vanishes while the geodesic term remains finite. 

The role of quantum geometry has been recently studied in magic-angle twisted bilayer graphene, where it is predicted to play an important role in superconductivity, light-matter coupling and nonlinear optical response \cite{ikeda2020high,orenstein2021topology,topp2021light,ahn2021riemannian,wu2020quantum}. We thus expect such moir\'e materials to be an ideal platform to explore geodesic dynamics. For the case of MATBG which has quasi-flat bands, the geodesic term should be the dominant longitudinal velocity. First, we estimate the minimum ($E_{\mathrm{min}}$) under which the geodesic response should be observable. We can define $E_{\mathrm{min}}$ to be such that the geodesic response is equal to the zeroth-order response i.e., the renormalized Dirac velocity $\tilde{v}_D$. For a sublattice gap $\Delta$ = 2 meV, we obtain $E_{\mathrm{min}}$ $\sim$ 55 mV/$\mu$m which is an experimentally reasonable value. The maximum electric field $E_{\mathrm{max}}$ must obey the perturbative limit of our derivation, 
\begin{equation*}
a \frac{e E_{\mathrm{max}}}{\hbar} \ll \Delta,    
\end{equation*}
where $a$ = 2.46 \(\text{\AA}\) is the lattice constant of monolayer graphene. Thus, we take $E_{\mathrm{min}}$ = 1 V/$\mu$m, which is a modest value for experiments. The geodesic for this value of electric field is shown in Fig.~3d and is $\sim$ $v_D$/40 which should be experimentally detectable.

\subsection{EFE equation}
The simplest way to create a mixed state which leads to a momentum-space entropic field is to consider thermal states over a set of bands. This state corresponds to replacing $p_n(\mathbf{k})$ in Eq.~\ref{densitymatrix} by $e^{-\beta \mathcal{E}_n(\mathbf{k})}/\mathcal{Z}$. In this way, the entropy varies in the momentum-space due to the finite dispersion of the bands. Our results imply that momentum-space wavepackets created from thermal states will experience an additional entropic (momentum-space) force. The modified Christoffel symbol of Eq.~\ref{connection} for the case of thermal mixed state with $k_B$ set to 1 becomes,
\begin{equation}
    \bar{\Gamma}_{ijl} - \frac{\beta}{2}\bar{g}_{ij}\left( \partial_{k^l}\langle \mathcal{E}({\mathbf{k}}) \rangle - \langle \partial_{k^l}\mathcal{E}({\mathbf{k}}) \rangle \right),
\end{equation}
where $\langle \ldots \rangle$ stands for the thermal average. The term in parenthesis can be rewritten in terms of group velocity $v_l = \partial_{k^l} \mathcal{E}$ as,
\begin{equation}
   \bar{\Gamma}_{ijl} + \frac{\beta^2}{2}\bar{g}_{ij}\left( \langle \mathcal{E} v_l \rangle - \langle \mathcal{E} \rangle \langle v_l \rangle \right).
   \label{thermalforce}
\end{equation}

As $T \rightarrow 0$, $\beta \rightarrow \infty$ but the covariance term goes to zero exponentially fast, making sure that correction to geodesic response vanishes. 

A possible platform to experimentally observe such an entropic force is the microcavity exciton-polariton system, wherein the quantum metric has been recently measured \cite{gianfrate2020measurement}. The wavepacket dynamics can be directly visualized because of the optically active nature of exicton-polaritons. In addition, they offer exquisite control of the band dispersion through the tuning of light-matter coupling and resonance condition between exciton and cavity modes, which can be used to tune the term in parenthesis of Eq.~\ref{thermalforce}. Non-resonant pumping of exciton-polaritons can be used to generate a quasi-thermal distribution over momentum states. Moreover, the tunability of the band gap between upper and lower polariton branches can be a knob to change the thermal distribution and vary the size of the entropic force. Due to the finite mass of photons in the cavity, a substantial part of the polariton dispersion lies within the light cone and can be probed optically.




\section{Conclusion}{\label{sec:Conclusion}}

In this work, we have highlighted the duality between the momentum-space equation
for the velocity of a Bloch electron and the position-space equation for the Lorentz force in curved space.
While the role of the metric in the second-order response of a Bloch electron to static electric and magnetic fields has been shown previously, the geodesic nature of these expressions has not previously been explored~\cite{gao2014field,gao2015geometrical,gao2019nonreciprocal,kozii2021intrinsic}.

Our results offer an intuitive picture for the role of quantum metric in the semiclassical dynamics of Bloch electrons in terms of momentum-space dual of the geodesic equation. Thus, the quantum metric can be thought to realize a momentum-space gravity. This analogy with the momentum-space gravity can be exploited to control charge carrier dynamics in solids, offering another knob in addition to the Berry curvature. Finally, our results which are derived for Bloch electrons, should also be relevant to other physical systems such as ultracold atoms in optical lattices for realizing synthetic gravity. 

As a future direction, extending our analysis beyond the two-band approximation might expose other geometric invariants of the interband Berry connection. Another extension of our work could be aimed at understanding the behavior in the limit of vanishing band gap, as in the case of graphene or Weyl semimetals. In addition, the momentum-space analogy can be applied in the study of nonlinear response to optical fields.

In addition to extending the duality present in the semiclassical formulation of electron dynamics, we have investigated the momentum-space dual of the EFE, the other fundamental equation of general relativity. We have shown that the presence of nonzero entropy of mixed states can lead to the emergence of a stress-energy tensor in the momentum space. Under the assumptions of small entropy, we have quantified the role of entropy in the creation of momentum-space stress-energy, and have shown that the stress-energy is in part due to the Laplacian of the entropy, just as the real-space stress-energy is due to the Laplacian of the gravitational potential in the Newtonian limit of general relativity~\cite{misner1973gravitation}.
We can thus view our expression \eqref{stressenergy} for stress-energy in the limit of low entropy as the analog of the Newtonian limit of general relativity. The momentum-space analog of the gravitational potential is the von Neumann entropy of a mixed quantum state, which is reminiscent of previous speculations on a potentially deep relationship between thermodynamic entropy and gravity~\cite{verlinde2011origin}. The von Neumann entropy, which enters our analysis, makes the connection to momentum-space gravity more general than the special case of thermodynamic entropy. As the von Neumann entropy in our case can be thought to result from the severing of quantum entanglement, our results appear  similar in spirit to the ongoing efforts aimed at showing quantum entanglement as the source of real space-time \cite{van2010building,maldacena2013cool}. 

A natural direction for future studies is to examine the case of arbitrarily large entropy. The connection with quantum entanglement is particularly interesting to explore. For example, one can gain further insight by studying the quantum evolution of a simple bipartite system,  which is interrupted by measurements. The unitary part of the dynamics will create entanglement which will then be severed by measurements to create entropy. A quantitative relation between entropy creation and changes in the geometry experienced by subsystems can further test the connections between quantum entanglement, quantum geometry and gravity.

\section*{Acknowledgements}
We acknowledge fruitful discussions with D. Xiao, Y. Gao, R. Cheng, H. Price, T. Ozawa, I. Carusotto and T. Chervy. We also thank W. Li for help with the figures. T.~B.~S. acknowledges funding from the Woodruff Fellowship of Emory University. L.~P. acknowledges funding from the Women in Natural Science Fellowship of Emory University. A.~S. acknowledges funding from the NSF EFRI program, grant no. EFMA-1741691, and from NSF DMR award no. 1905809. 

\appendix
\section{Useful relations}{\label{Appendix A}}

\subsection*{1. Definitions and identities}
Before deriving the equations presented in the main text, we first note some useful preliminaries.

The Berry connection for a given band ($n$) is defined as~\cite{xiao2010berry}:
    \begin{equation}
        \mathcal{A}_{n}(\mathbf{k}) = i \bra{u_{n}(\mathbf{k})} \nabla_{\mathbf{k}} \ket{u_{n}(\mathbf{k})}.
    \end{equation}
The interband Berry connection between bands $m$ and $n$ takes a similar form:
    \begin{equation}
        \mathcal{A}_{mn}(\mathbf{k}) = i \bra{u_{m}(\mathbf{k})} \nabla_{\mathbf{k}} \ket{u_{n}(\mathbf{k})},
    \end{equation}
and for $m=n$, the interband Berry connection reduces to the  (intraband) Berry connection for a single band.
    
The Berry curvature of a band is defined as the curl of the Berry connection:
    \begin{equation}
        \label{1.3}
        \Omega_{n} (\mathbf{k}) = \nabla_{\mathbf{k}} \cross \mathcal{A}_{n}(\mathbf{k}) = -2 Im[\braket{\partial_{k^{i}}u_{n}}{\partial_{k^{j}}u_{n}}],
    \end{equation}
and the quantum metric is defined via the Berry connection and derivatives of the Bloch function~\cite{srivastava2015signatures}:
    \begin{equation}
        \label{1.4}
        g_{ij}(\mathbf{k}) = Re[\braket{\partial_{k^{i}}u_{n}(\mathbf{k})}{\partial_{k^{j}}u_{n}(\mathbf{k})}] - \mathcal{A}_{i}(\mathbf{k}) \mathcal{A}_{j}(\mathbf{k}).
    \end{equation}

\subsection*{2. Useful relations using Bloch bands}

\noindent For any number of bands, the following relation between the Bloch function and the quantum metric holds:

   \begin{equation}
    \label{1.5}
    1 - |\braket{u_{n}}{\Tilde{u_{n}}}|^{2} \approx g_{ij} A^{i} A^{j},
\end{equation}
where the perturbed bloch function $ |\Tilde{u_{n}} \rangle$ is defined as $|u_{n}(\mathbf{k} - \frac{e}{\bar{h}} \mathbf{A}) \rangle$, the unperturbed Bloch function $ |u_{n} \rangle$  is $  |u_{n}(\mathbf{k}) \rangle$, and $\approx$ denotes equality up to second order in the external field, $\mathbf{A}$. 

We also note the relation between the interband Berry connection and the quantum metric and Berry curvature:
 
 \begin{equation}  
    \label{1.7}
    \sum_{n \neq 0} \mathcal{A}_{0n} \mathcal{A}_{n0} = g_{ij,0} - \frac{i}{2} \Omega_{ij,0}.
\end{equation}

\noindent In a 2 band model, \ref{1.7} reduces to: 
\begin{equation}
    \label{1.6}
    |\braket{u_{0}}{\Tilde{u_{1}}}|^{2} = g_{ij} A^{i} A^{j} .
\end{equation}
To derive equation \ref{1.7}, we note the relation between the interband Berry connection and the projection operator:
$$
\sum_{n \neq 0} \mathcal{A}_{0n,i} \mathcal{A}_{n0,j} = - \sum_{n \neq 0}\bra{u_{0}} \partial_{k^{i}} \ket{u_{n}} \bra{u_{n}} \partial_{k^{j}} \ket{u_{0}}.
$$
We can re-express the right-hand side of the above equation as
$$
\begin{gathered}
\sum_{n \neq 0} \mathcal{A}_{0n,i} \mathcal{A}_{n0,j} = \\  -\bra{u_{0}} \partial_{k^{i}}\partial_{k^{j}} \ket{u_{0}} + \bra{u_{0}} \partial_{k^{i}} \ket{u_{0}} \bra{u_{0}} \partial_{k^{j}} \ket{u_{0}},
\end{gathered}
$$
which is equivalent to
$$
\begin{gathered}
\sum_{n \neq 0} \mathcal{A}_{0n,i} \mathcal{A}_{n0,j} = \\    Re \big[\braket{\partial_{k^{i}}u_{0}}{\partial_{k^{j}}u_{0}} \big]  - \mathcal{A}_{00,i}\mathcal{A}_{00,j} + i Im \big[\braket{\partial_{k^{i}}u_{0}}{\partial_{k^{j}}u_{0}} \big].
\end{gathered}
$$

\vspace{0.07 in}
\subsection*{3. Perturbative relations}

   Up to second order in the applied vector potential, $A$, the perturbed Bloch function can be written as 
    \begin{align}
        \ket{\Tilde{u_{n}}} = \ket{u_{n}} + \ket{\partial_{k^{i}}u_{n}} A^{i} + \frac{1}{2} \ket{\partial_{k^{j}}\partial_{k^{i}}u_{n}} A^{i}A^{j}.
    \end{align}

\noindent Using the above expansion, we can write second-order expressions for the outer products of Bloch functions as,

\begin{widetext}

    \begin{equation}
    \label{1.9}
        \begin{split}
            \ket{\Tilde{u_{0}}}\bra{\Tilde{u_{0}}} & =  \ket{u_{0}} \bra{u_{0}} + A^{i} \Big[ \ket{\partial_{k^{i}}u_{0}} \bra{u_{0}} + \ket{u_{0}}\bra{\partial_{k^{i}}u_{0}} \Big] + \\ & \frac{1}{2} \Big[\ket{\partial_{k^{i}} \partial_{k^{j}}u_{0}} \bra{u_{0}} + \ket{u_{0}}\bra{\partial_{k^{j}}\partial_{k^{i}}u_{0}}  + 2 \ket{\partial_{k^{i}}u_{0}}\bra{\partial_{k^{j}}u_{0}}\Big]A^{i}A^{j},
        \end{split}
    \end{equation}
    
    \begin{equation}
    \label{1.10}
        \begin{split}
            \ket{\Tilde{u_{1}}}\bra{\Tilde{u_{1}}} & =  \ket{u_{1}} \bra{u_{1}} + A^{i} \Big[ \ket{\partial_{k^{i}}u_{1}} \bra{u_{1}} + \ket{u_{1}}\bra{\partial_{k^{i}}u_{1}} \Big] + \\ & \frac{1}{2} \Big[\ket{\partial_{k^{i}} \partial_{k^{j}}u_{1}} \bra{u_{0}} + \ket{u_{0}}\bra{\partial_{k^{j}}\partial_{k^{i}}u_{1}}  + 2 \ket{\partial_{k^{i}}u_{1}}\bra{\partial_{k^{j}}u_{1}}\Big]A^{i}A^{j}.
        \end{split}
    \end{equation}
\end{widetext}

\section{Derivation of the momentum-space geodesic term}{\label{Appendix B}}

\subsection*{1. Determining the perturbed Hamiltonian}

Consider a 2 band model for which band indices are $0$ and $1$. The unperturbed Hamiltonian in its eigenbasis can be expressed as
\begin{equation}
    H(\mathbf{k}) = \sum_{n} \bra{u_{n}}H \ket{u_{n}} \ket{u_{n}} \bra{u_{n}} = \sum_{n} \epsilon_{n} \ket{u_{n}} \bra{u_{n}}
\end{equation}
\noindent In the presence of an electromagnetic field defined by the potential $\mathbf{A}(\mathbf{x},t)$, the Hamiltonian becomes $\Tilde{H} = H(\mathbf{k} - \frac{e}{\bar{h}} \mathbf{A}) $. The perturbed Hamiltonian in its eigen basis is then:

\begin{equation}
    \Tilde{H} = \sum_{n} \bra{\Tilde{u_{n}}} \Tilde{H} \ket{\Tilde{u_{n}}} \ket{\Tilde{u_{n}}} \bra{\Tilde{u_{n}}} = \sum_{n} \Tilde{\epsilon_{n}} \ket{\Tilde{u_{n}}} \bra{\Tilde{u_{n}}}.
\end{equation}

In the unperturbed basis, the perturbed Hamiltonian has the following matrix elements:

\begin{align}
     \bra{u_{m}} \Tilde{H} \ket{u_{n}} =  \Tilde{\epsilon_{0}} \braket{u_{m}}{\Tilde{u_{0}}} \braket{\Tilde{u_{0}}}{u_{n}} + \Tilde{\epsilon_{1}} \braket{u_{m}}{\Tilde{u_{1}}} \braket{\Tilde{u_{1}}}{u_{n}}.
\label{perturbed_hamiltonian_components}
\end{align}

For the diagonal element $H_{00}$, \eqref{perturbed_hamiltonian_components} yields

$$
 \Tilde{H_{00}} = \Tilde{\epsilon_{0}}|\braket{u_{0}}{\Tilde{u_{0}}}|^{2}  +  \Tilde{\epsilon_{1}}  |\braket{u_{0}}{\Tilde{u_{1}}}|^{2}.
$$
\noindent Using equation \ref{1.5} and \ref{1.6}, $H_{00}$ can be rewritten as
$$
    \Tilde{H_{00}} = \Tilde{\epsilon_{0}}  + \big( \Tilde{\epsilon_{1}} - \Tilde{\epsilon_{0}} \big) g_{ij} A^{i} A^{j}.
$$
\noindent We find a similar expression for  $H_{11}$:
$$
    \Tilde{H_{11}} = \Tilde{\epsilon_{1}}  + \big( \Tilde{\epsilon_{0}} - \Tilde{\epsilon_{1}} \big) g_{ij} A^{i} A^{j}.
$$

For the off-diagonal elements, we have
$$
\Tilde{H}_{01} =  \Tilde{\epsilon_{0}} \braket{u_{0}}{\Tilde{u_{0}}} \braket{\Tilde{u_{0}}}{u_{1}} + \Tilde{\epsilon_{1}} \braket{u_{0}}{\Tilde{u_{1}}} \braket{\Tilde{u_{1}}}{u_{1}}
$$
and 
$$
\Tilde{H}_{10} =  \Tilde{\epsilon_{1}} \braket{u_{1}}{\Tilde{u_{1}}} \braket{\Tilde{u_{1}}}{u_{0}} + \Tilde{\epsilon_{0}} \braket{u_{1}}{\Tilde{u_{0}}} \braket{\Tilde{u_{0}}}{u_{0}}.
$$

To simplify our analysis of the off-diagonal elements, we will assume that the bands are flat, such that $\Tilde{\epsilon_{0}} = \epsilon_{0}$.
On substituting for the correction in state from equation \ref{1.9} and \ref{1.10}, we find an expression for the off-diagonal elements of the Hamiltonian in the unperturbed basis:

\begin{equation}
    \begin{gathered}
        \Tilde{H}_{mn} =  \Big[ i (\Tilde{\epsilon_{m}}-\Tilde{\epsilon_{n}}) + \sum_{l=0,1} \Tilde{\epsilon}_{l} \mathcal{\mathbf{A}}_{ll} \cdot \mathbf{A} \Big] \mathcal{\mathbf{A}}_{mn} \cdot \mathbf{A} + \\ \frac{A^{i}A^{j}}{2} \big(\Tilde{\epsilon_{m}} \braket{\partial_{k^{j}}\partial_{k^{i}}u_{m}}{u_{n}} + \Tilde{\epsilon_{n}} \braket{u_{m}}{\partial_{k^{j}}\partial_{k^{i}}u_{n}} \big)  \hspace{0.05 in} \\
        \text{ for $ m \neq n$}.
    \end{gathered}
\end{equation}

\subsection*{2. Time-dependent corrections }
\noindent Let $\mathbf{A}(\mathbf{x},t) = -\frac{\mathbf{E_{0}}}{\omega} \sin{(\omega t)}$. Beginning with the ansatz $\ket{\psi(t)} = C_{0}(t)e^{-\epsilon_0 t} \ket{u_{0}} + C_{1}(t)e^{-\epsilon_1 t} \ket{u_{1}}$ and $C_{0}(0)=1$, we can apply the standard methods of time-dependent perturbation theory to find corrections to the wave function and the band geometry.  

The wave function coefficients are found via the following equations:

\begin{widetext}
    \begin{equation}
        i \frac{\partial C_{0}}{\partial t}(t )  =  \Big[ i (\epsilon_{0}-\epsilon_{1}) \mathcal{A}_{01,i} A^{i} - \frac{(\epsilon_{0}+\epsilon_{1})}{2} \mathcal{A}_{10,i} \mathcal{A}_{01,j} A^{i}A^{j} \Big]e^{-i\Delta t} C_{1}(t) -  \Big[\big( \epsilon_{1} - \epsilon_{0} \big) g_{ij} A^{i} A^{j} \Big] C_{0}(t)
    \end{equation}

and

    \begin{equation}
        i \frac{\partial C_{1}}{\partial t}(t )  =  \Big[ i (\epsilon_{1}-\epsilon_{0}) \mathcal{A}_{10,i} A^{i} - \frac{(\epsilon_{1}+\epsilon_{0})}{2} \mathcal{A}_{01,i} \mathcal{A}_{10,j} A^{i}A^{j} \Big]e^{i\Delta t} C_{0}(t) -  \Big[\big( \epsilon_{0} - \epsilon_{1} \big) g_{ij} A^{i} A^{j} \Big] C_{1}(t),
    \end{equation}
\end{widetext}

where the coefficients to known order are used on the right-hand side to compute coefficients of the next-highest order.

The zeroth order coefficients are simply $C^0_0(t) = 1$ and $C^0_1(t) = 0$.
The equations for the first order corrections are found to be:

\begin{equation}
i\partial_t C_{0}^{\prime}(t) = 0
\end{equation}
and
$$
\begin{gathered}
i \frac{\partial C_{1}^{\prime}}{\partial t}(t) = -i \Delta \mathcal{A}_{10} \frac{\mathbf{E_{0}}}{\omega} \sin{(\omega t)} e^{i \Delta t},
\end{gathered}
$$
where $\Delta = \epsilon_{1}-\epsilon_{0}$.

The resulting first-order correction to $C_1(t)$ is
\begin{equation}
    C_{1}^{\prime}(t) = \frac{\Delta}{2  \omega} \mathcal{A}_{10} \mathbf{E_{0}} \Big[ \frac{e^{i (\omega + \Delta)t}-1}{\omega + \Delta} + \frac{e^{-i (\omega - \Delta)t}-1}{\omega - \Delta} \Big].
\label{first_order_C1}
\end{equation}

\begin{widetext}

Now turning to second order, the correction to $C_0(t)$ is governed by the following equation:
$$
i \frac{\partial C_{0}^{\prime \prime}}{\partial t}(t) =  \Big[  i (\epsilon_{0}-\epsilon_{1}) \mathcal{A}_{01,i} A^{i} \Big]e^{-i\Delta t} C_{1}^{\prime}(t) +  \Big[\big( \epsilon_{1} - \epsilon_{0} \big) g_{ij} A^{i} A^{j} \Big] C_{0}^{0}(t),
$$
which can be re-expressed as:

$$
i \frac{\partial C_{0}^{\prime \prime}}{\partial t}(t) =  -\Big[  i \Delta \mathcal{A}_{01,i} \frac{E_{0,i}}{\omega} \sin{(\omega t)} \Big] e^{-i\Delta t}C_{1}^{\prime}(t) +   \Big[\Delta \big( g_{ij}-\frac{1}{2} \Omega_{ij}\big) E_{0}^{i} E_{0}^{j} \big(\frac{\sin{(\omega t)}}{\omega} \big)^{2} \Big].
$$

\noindent Note that the $\Omega_{ij}$ terms are anti symmetric in $i$ and $j$ while $E_{0}^{i}E_{0}^{j}$ is symmetric in $i$ and $j$. 
This expression can thus be further simplified to:

$$
i \frac{\partial C_{0}^{\prime \prime}}{\partial t}(t) =  - i\Delta^2 g_{ij} E_{0}^{i} E_{0}^{j}  \big(\frac{ \sin{(\omega t)}}{2\omega^2} \big)\Big[ \frac{e^{i \omega t}-e^{-i \Delta t}}{\omega + \Delta} + \frac{e^{-i \omega t}-e^{-i \Delta t}}{\omega - \Delta} \Big]   +  \Delta  g_{ij} E_{0}^{i} E_{0}^{j} \big(\frac{\sin{(\omega t)}}{\omega} \big)^{2}.
$$
The result for the second order correction is:
\begin{equation}
    C_{0}^{\prime \prime}(t) = \alpha(t) g_{ij} E_{0}^{i} E_{0}^{j},
\label{C_0_second_rder}
\end{equation}
\noindent where $\alpha(t)$ can be defined in terms of its real and imaginary components:

$$ 
Re[\alpha(t) ]= 
\frac{\Delta^2}{(\omega^2 - \Delta^2)^2}\left( 1 - \cos(\Delta t)\cos(\omega t) \right) - 
\frac{\Delta^3}{\omega(\omega^2 - \Delta^2)^2}\left(  \sin(\Delta t)\sin(\omega t) \right) + 
\frac{\Delta^2}{4\omega^2(\omega^2 - \Delta^2)}\left( \cos(2 \omega t) -1 \right), 
$$

$$
Im[\alpha(t) ]= 
\frac{\Delta^2}{(\omega^2 - \Delta^2)^2}\left( \sin(\Delta t)\cos(\omega t)\right) - 
\frac{\Delta^3}{\omega(\omega^2 - \Delta^2)^2}\left( \cos(\Delta t)\sin(\omega t)\right) - 
\frac{\Delta(\omega^2 - 2 \Delta^2)\left( 2\omega t- \sin(2 \omega t )\right)}{4(\omega^2 - \Delta^2)\omega^3}.
$$

We find a similar relation governing the second order correction to $C_1(t)$:

$$
i \frac{\partial C_{1}^{\prime \prime}}{\partial t}(t )  = \Big[ \epsilon_{0} \mathcal{A}_{00,i} \mathcal{A}_{10,j} + \epsilon_{1} \mathcal{A}_{11,i} \mathcal{A}_{10,j} + \frac{\epsilon_{0}}{2} \braket{u_{1}}{\partial_{k^{i}}\partial_{k^{j}}u_{0}} + \frac{\epsilon_{1}}{2} \braket{\partial_{k^{i}}\partial_{k^{j}}u_{1}}{u_{0}} \Big] E_{0}^{i}E_{0}^{j} \frac{[\sin{(\omega t)}]^{2}}{\omega^{2}}
$$

which has the following solution:

\begin{equation}
   C_{1}^{\prime \prime}(t) =  K_{ij} \frac{E_{0}^{i}E_{0}^{j}}{\omega^{2}} \Big(\frac{i(e^{i(\Delta + 2\omega) t} -1)}{4 (\Delta + 2 \omega)} + \frac{i(e^{i(\Delta - 2\omega) t} -1)}{4 (\Delta - 2 \omega)} - \frac{i(e^{i\Delta t} -1)}{2 \Delta}\Big),
\label{C_1_second_order}
\end{equation}

where $K_{ij} = -i \Big[ \epsilon_{0} \mathcal{A}_{00,i} \mathcal{A}_{10,j} + \epsilon_{1} \mathcal{A}_{11,i} \mathcal{A}_{10,j} + \frac{\epsilon_{0}}{2} \braket{u_{1}}{\partial_{k^{i}}\partial_{k^{j}}u_{0}} + \frac{\epsilon_{1}}{2} \braket{\partial_{k^{i}}\partial_{k^{j}}u_{1}}{u_{0}} \Big] $ is a time-independent quantity.

\end{widetext}

\subsection*{3. Corrections to the Berry connection}
\noindent We represent the corrected Berry connection (up to second order) as $\Tilde{\mathcal{A}} = \mathcal{A}+a^{\prime}+a^{\prime \prime}$.

The first order correction is defined as:
$$
a^{\prime} = i \Big[ \bra{u_{0}} \mathbf{\nabla_{k}} \ket{u_{1}} e^{-i \Delta t} C_{1}^{\prime}(t) + \bra{u_{1}} \mathbf{\nabla_{k}} \ket{u_{0}} e^{i \Delta t} C_{1}^{\prime *}(t) \Big],
$$

which simplifies to:
$$
a^{\prime}_{i} = \mathcal{A}_{01,i}  e^{-i \Delta t}  C_{1}^{\prime}(t) + \mathcal{A}_{10,i}  e^{i \Delta t}  C_{1}^{\prime *}(t).
$$
Using expression \ref{first_order_C1} for $C_{1}^{\prime}(t)$, the above relation becomes:

$$
\begin{gathered}
    a^{\prime}_{i} = \frac{\Delta}{2 \omega} \mathcal{A}_{01,i} \mathcal{A}_{10,j}E_{0}^{j} \Big[\frac{e^{i \omega t}- e^{-i \Delta t}}{\omega + \Delta} + \frac{e^{-i \omega t}-e^{-i \Delta t}}{\omega - \Delta}  \Big] + \\ \frac{\Delta}{2 \omega} \mathcal{A}_{01,j} \mathcal{A}_{10,i}E_{0}^{j} \Big[\frac{e^{-i \omega t}- e^{i \Delta t}}{\omega + \Delta} + \frac{e^{i \omega t}-e^{i \Delta t}}{\omega - \Delta}  \Big].
\end{gathered}
$$

We then use \ref{1.7} to re-express $a^{\prime}$ in terms of the Berry curvature and quantum metric:

\begin{equation}
    a^{\prime}_{i} = \frac{\Delta}{\omega} E_{0}^{j} \Big[ g_{i j,0}  S^{\omega}(t) +  \frac{\Omega_{i j,0}}{2} A^{\omega}(t) \Big],
\end{equation} 
where $S^{\omega}(t) = \frac{2 \cos{\omega t} - 2 \cos{\Delta t}}{\omega + \Delta} + \frac{2 \cos{\omega t} - 2 \cos{\Delta t}}{\omega - \Delta} $ and $A^{\omega} (t) = \frac{2 i \sin{\omega t} + 2 i \sin{\Delta t}}{\omega + \Delta} + \frac{-2 i \sin{\omega t} + 2 i \sin{\Delta t}}{\omega - \Delta}$.

In order to achieve a steady-state, a relaxation time, $\tau$ due to scattering with impurities, phonons, etc is required. Using the zero frequency limit ($\omega \to 0$) and the long time approximation ($t \gg \tau$, such that all oscillatory terms drop out), we find the following steady-state expression for $a^{\prime}$:

\begin{equation}
    \label{1.27}
    a^{\prime}_{i} = -\frac{2}{\Delta} g_{ij}E_{0}^{j}, 
\end{equation}
which agrees with previous results found via time-independent perturbation theory~\cite{gao2014field}.

We can extend our correction of the Berry connection to second order using our second-order corrections to the wave function.  As we are interested in the observable response generated by the connection, we consider the gauge invariant portion (``positional shift") of the second order correction, which we denote as $a^{\prime \prime}(t)$.
$a^{\prime \prime}(t)$ can be expressed as:

\begin{widetext}
$$
a^{\prime \prime}(t) =  i \Big[ \bra{u_{0}} \mathbf{\nabla_{k}} \ket{u_{1}} e^{-i \Delta t} \big( C_{0}^{*}C_{1}^{\prime \prime} + C_{0}^{\prime *}C_{1}^{\prime } + C_{0}^{\prime \prime *}C_{1} \big)\Big] +   i \Big[\bra{u_{1}} \mathbf{\nabla_{k}} \ket{u_{0}} e^{i \Delta t} \big( C_{1}^{*}C_{01}^{\prime \prime} + C_{1}^{\prime *}C_{0}^{\prime } + C_{1}^{\prime \prime *}C_{0}\big)   \Big],
$$

which simplifies to:

$$
a^{\prime \prime}(t) =  i \bra{u_{0}} \mathbf{\nabla_{k}} \ket{u_{1}} e^{-i \Delta t} C_{1}^{\prime \prime}(t) + i \bra{u_{1}} \mathbf{\nabla_{k}} \ket{u_{0}} e^{i \Delta t} C_{1}^{\prime \prime *}(t).
$$

Using our expression \ref{C_1_second_order} for $C_{1}^{\prime \prime}(t)$, the second order correction to the Berry connection can be written as

\begin{equation}
\begin{split}
    a^{\prime \prime}_{l}(t) & =  - \mathcal{A}_{01,l}  K_{ij} \Big(\frac{(e^{i( 2\omega) t} -e^{-i \Delta t})}{4 (\Delta + 2 \omega)} + \frac{(e^{-i(  2\omega) t} -e^{-i \Delta t})}{4 (\Delta - 2 \omega)} - \frac{(1 -e^{-i \Delta t})}{2 \Delta}\Big) \frac{E_{0}^{i}E_{0}^{j}}{\omega^{2}} + \\ &  \mathcal{A}_{10,l}  K_{ij}^{*} \Big(\frac{(e^{-i( 2\omega) t} -e^{i \Delta t})}{4 (\Delta + 2 \omega)} + \frac{(e^{i( 2\omega) t} -e^{i \Delta t})}{4 (\Delta - 2 \omega)} - \frac{(1 -e^{i \Delta t})}{2 \Delta}\Big) \frac{E_{0}^{i}E_{0}^{j}}{\omega^{2}},
\end{split} 
\end{equation}

which simplifies considerably in the low frequency limit:

\begin{equation}
\begin{gathered}
    a^{\prime \prime}_{l}(t) =  - \mathcal{A}_{01,l}  K_{ij}E_{0}^{i}E_{0}^{j} \Big( \frac{(1 -e^{-i \Delta t})}{2 \Delta^3}\Big)  + \mathcal{A}_{10,l}  K_{ij}^{*}E_{0}^{i}E_{0}^{j} \Big( \frac{(1 -e^{i \Delta t})}{2 \Delta^3}\Big) .
\label{low_freq_a''}
\end{gathered} 
\end{equation}

\end{widetext}

To reach the steady-state associated with long times, we average over the oscillatory terms, leaving only the time-independent portion of Eq.~\ref{low_freq_a''}.

Now that we have derived the corrections $a'$ and $a''$, we can consider the electron response associated with these quantities.  In addition to the first-order correction of the Berry curvature, defined as $\nabla_{\mathbf{k}} \times \mathbf{a}'$, the temporal derivatives of $a^{\prime}(t)$ and $a^{\prime \prime}(t)$ can also contribute to the response.
The time derivative of  $a^{\prime}(t)$ is found to be:
\begin{equation}
\begin{gathered}
    \partial_{t} a_{i}^{\prime}(t) = -\frac{2 \Delta}{\omega^{2}-\Delta^{2}} g_{ij}E_{0}^{j} \big(\omega \sin{(\omega t)}- \Delta \sin{(\Delta t)}  \big) 
\end{gathered} 
\end{equation}
While nonzero for AC fields, this term vanishes in the zero frequency limit when we apply the long time approximation and average over oscillating terms.

Likewise, the partial derivative in time of $a^{\prime \prime}(t)$ yields a second-order response, which takes a simple form in the zero-frequency limit:

\begin{equation}
\begin{gathered}
    \partial_t a^{\prime \prime}_{l}(t) =  - \mathcal{A}_{01,l}  K_{ij}E_{0}^{i}E_{0}^{j} \Big( \frac{( i \Delta e^{-i \Delta t})}{2 \Delta^3}\Big)  + c.c. ,
\label{low_freq_a''}
\end{gathered} 
\end{equation}
where $c.c.$ denotes the complex conjugate of the first term.
\noindent Upon taking the long time approximation, the oscillatory terms vanish (average to zero) and $\partial_{t} a^{\prime \prime} \longrightarrow 0$.

\subsection*{4. Corrections in Energy}

The energy of a time-dependent state in the presence of an external field can be written as:
$$\bra{\Tilde{u_{0}}} \Tilde{H} \ket{\Tilde{u_{0}}},$$
where both the wave function and Hamiltonian must be corrected for the presence of the perturbation.

Expanding the corrected Hamiltonian $\Tilde{H}$ as $H_0 + H^{\prime} + H^{\prime \prime}$, we keep terms of necessary order in the wave function coefficients and find the following expression for energy:

$$
\Tilde{\epsilon} = |C_{0}(t)|^{2} \epsilon_{0} + |C_{1}(t)|^{2} \epsilon_{1} + 2Re[H^{\prime}_{0 1} C_1^{\prime}e^{-i\Delta t}] + H^{\prime \prime}_{00},
$$
where $|C_{0}(t)|^{2} = 1 -|C_{1}(t)|^{2} $ (up to second order) due to the constraint of normalization.  Writing all quantities in terms of the quantum metric, we find
that the corrected energy (to second order) can be written as:

\begin{equation}
    \Tilde{\epsilon} = \epsilon_{0}+ \Delta \beta(t)g_{ij} E_{0}^{i}E_{0}^{j},
\end{equation}
where $\beta$ is defined as:
\begin{widetext}
\begin{equation}
\begin{gathered}
\beta(t) =   \frac{\Delta^2}{4\omega^2}\left(\frac{\cos((\omega + \Delta) t) - 1}{ \omega + \Delta}  + \frac{\cos((\omega - \Delta) t) - 1}{ \omega - \Delta}\right)^2 +  \frac{\Delta^2}{4\omega^2}\left(\frac{\sin((\omega + \Delta) t) }{ \omega + \Delta}  - \frac{\sin((\omega - \Delta) t) }{ \omega - \Delta}\right)^2 + \\
\frac{\Delta\sin(\omega t)}{\omega^2}\left( \frac{\sin(\omega t) + \sin(\Delta t)}{\omega + \Delta} + \frac{\sin(\Delta t) - \sin(\omega t)}{\omega - \Delta} \right) + 
\frac{\sin^2(\omega t)}{\omega^2}. 
\end{gathered}
\end{equation}
\end{widetext}

The corrected energy (to second order) can then be written as:
\begin{equation}
    \Tilde{\epsilon} = \epsilon_{0}+ \Delta \beta(t)g_{ij} E_{0}^{i}E_{0}^{j}.
\end{equation}

\noindent The zero frequency, steady-state limit of $\beta(t)$ (after averaging over oscillations) is
\begin{equation}
    \beta(t) \longrightarrow \frac{2}{\Delta^{2}}.
\end{equation}

\subsection*{5. Steady-state analysis in the dipole gauge}
We begin with the dipole Hamiltonian:

\begin{equation}
    H(\mathbf{k}) = H_{0}(\mathbf{k}) + \mathbf{E} \cdot \mathbf{r},
\end{equation}
\noindent where $\mathbf{r}$ is the position operator with elements $\mathbf{r}_{mn} =  \mathcal{A}_{mn}$ for $m \neq n$.  We set the diagonal elements of the dipole correction to zero, as these terms only shift energies and do not generate transitions between bands.
These diagonal elements will be irrelevant for the following analysis.

\noindent The correction to the Hamiltonian can be expressed in matrix form as
\begin{equation}
    H^{\prime} = \begin{bmatrix}
    0 & \mathbf{E} \cdot \mathbf{\mathcal{A}}_{01}(\mathbf{k}) \\
    \mathbf{E} \cdot \mathbf{\mathcal{A}}_{10}(\mathbf{k}) & 0 \\
\end{bmatrix}.
\end{equation}

Using $H^{\prime} $, we can again apply time-dependent perturbation theory to compute the wave function corrections for this static perturbation.
The resulting equations for the coefficients are:

$$
\begin{gathered}
i \partial_t C_0(t) =  \mathbf{E} \cdot \mathbf{\mathcal{A}_{01}} C_1(t) e^{-i \Delta t} \\
\end{gathered}
$$

and

$$
 i \partial_t C_1(t) =   \mathbf{E} \cdot \mathbf{\mathcal{A}_{10}} C_0(t) e^{i \Delta t}
$$
\noindent with initial conditions  $C_{0}(t = 0) = 1$ and $C_{1}(t=0) = 0$.

Using zeroth order coefficients of 1 and 0, the first order corrections are given by:
$$
 \partial_t C_0^{'}(t) = -i\mathbf{E} \cdot \mathbf{\mathcal{A}_{01}} C_1^{0}(t) e^{-i \Delta t} = 0
$$

and

$$
 \partial_t C_1^{'}(t) = -i\mathbf{E} \cdot \mathbf{\mathcal{A}_{10}} C_0^{0}(t) e^{i \Delta t} = -i\mathbf{E} \cdot \mathbf{\mathcal{A}_{10}} e^{i \Delta t}.
$$
The resulting coefficients (to first order) are $C_{0}(t) = 1 $ and $C_{1}(t) = \frac{1 - e^{i \Delta t}}{\Delta} E_{0}^{i} \mathcal{A}_{01,i}$.
We can then use the wave function corrections to correct the Berry connection to first order:
$$
a^{\prime} = i \Big[ \bra{u_{0}} \mathbf{\nabla_{k}} \ket{u_{1}} e^{-i \Delta t} C_{1}^{\prime}(t) + \bra{u_{1}} \mathbf{\nabla_{k}} \ket{u_{0}} e^{i \Delta t} C_{1}^{\prime *}(t) \Big],
$$
which results in an expression identical to \ref{1.27} at long times (after averaging over oscillations):

\begin{equation}
  a^{\prime}_{i} = -\frac{2}{\Delta} g_{ij}E_{0}^{j}.  
\end{equation}

We can also compute the second-order energy correction for a semiclassical wave packet:

$$\Tilde{\epsilon} =  \epsilon_{0} + |C_{1}(t)|^{2} \Delta,$$

which can be rewritten as
$$
\Tilde{\epsilon}  = \epsilon_{0} + g_{ij} E_{0}^{i} E_{0}^{j} \big( \frac{2 -e^{-i \Delta t} + e^{i \Delta t} }{\Delta} \big).
$$
Under the long time approximation (averaging over oscillations), the above equation reduces to the following steady-state expression:
\begin{equation}
   \Tilde{\epsilon} = \epsilon_{0} +  \frac{2}{\Delta} E_{0}^{i}E_{0}^{j} g_{ij}.
\end{equation}

\section{Derivation of Momentum-space EFE and Entropic force}{\label{Appendix C}}

\subsection{Calculating the metric and stress-energy tensor for mixed states}

The quantum metric for pure states is an Einstein metric in any dimension, and is thus a vacuum solution of the Einstein field equations~\cite{besse2007einstein}.   
 Here we show the that the Bures metric for mixed states need not be a vacuum solution, and can instead satisfy the Einstein field equations with a nonzero stress-energy tensor.
 Given that the two-dimensional case is trivial, with all metrics satisfying the vacuum equations with zero cosmological constant, we show this for a three dimensional momentum space.  
 
We consider the density matrix of a mixed state: 
\begin{equation}\rho = \sum_{n=0}^{N} p_n(\mathbf k ) | u_n(\mathbf k) \rangle  \langle u_n(\mathbf k ) |. \end{equation}

The difference between mixed state density matrices at nearby points in momentum space, $d \rho$, is found to be:
\begin{equation}  \langle u_i(\mathbf k ) | d\rho  | u_j(\mathbf k )\rangle =   \left( d \mathbf{k} \cdot \nabla_{\mathbf k} p_i  \right) \delta_{i j}    + i  ( p_i - p_j ) \mathcal{A}_{ij} \cdot d \mathbf{k}, 
\label{drho}
\end{equation}
where $ \mathcal{A}_{i j} $ is the interband Berry connection.

Using \eqref{drho} we can calculate the distance between mixed states as defined by the Bures metric~\cite{dittmann1999explicit}:
\begin{equation} \bar{g}_{i j}dk^i dk^j = \frac{1}{2} \sum_{j,k=0}^{N} \frac{|\langle j| d\rho | k\rangle  |^2}{p_j+p_k},
\end{equation}
which can be re-expressed as:

\begin{equation}
    \begin{gathered}
        \bar{g}_{i j}dk^i dk^j = \frac{1}{2} \sum_{j,k=0}^{N} \frac{|\mathbf{dk} \cdot \nabla_{\mathbf k} p_j  |^2}{2p_j} \delta_{j k} + \\ \frac{( p_j - p_k )^2}{ p_j + p_k } |\mathcal{A}_{jk} \cdot d \mathbf{k}|^2.    
    \end{gathered}
\end{equation}
Assuming the probabilities $p_n(\mathbf{k})$ change slowly in k-space (more precisely, $\nabla_\mathbf{k} \left( \mathrm{log}( p_n(\mathbf{k}))\right) \approx 0$), the above expression reduces to
 \begin{equation} \bar{g}_{i j}dk^i dk^j \approx  \frac{1}{2} \sum_{j,k=0}^{N}  \frac{( p_j - p_k )^2}{ p_j + p_k } |\mathcal{A}_{jk} \cdot d \mathbf{k}|^2.  
 \label{slowp_qgt}
 \end{equation}
First, we consider a thermal mixed state such that the Boltzmann weights are  $ p_n(\mathbf{k}) = e^{- \beta \mathcal{E}_n(\mathbf{k})}/ \mathcal{Z}(\mathbf{k}) $. Later, we will generalize our results to more general density matrices under the assumptions stated below. 
As we wish to focus on a single-band, say the ground state, we assume that $p_0 \gg p_i$ $\forall$ $i \neq 0$. For the thermal mixed state, this is equivalent to assuming a large energy gap such that all bands above the lowest band have 
very small probabilities. Eq.~\eqref{slowp_qgt} then reduces
 to

\begin{equation} 
\bar{g}_{i j}dk^i dk^j \approx \left(e^{-\beta \mathcal{E}_0}/\mathcal{Z} \right)  g_{ i j}dk^i dk^j, 
\label{mixedmetric}
\end{equation}
where we have once again re-expressed the outer product of the interband Berry connections as the pure state quantum metric.
Next, we show that the scale factor in front of the pure state metric depends on the entropy. Rewriting the scale factor as a conformal scale factor, $e^{2 f(\mathbf{k})}$, we see that,
\begin{eqnarray}
    f &=& -\frac{1}{2}\left( \beta \mathcal{E}_0(\mathbf{k}) + \mathrm{log}\mathcal{Z}(\mathbf{k}) \right) \\
    &\approx& -\frac{1}{2}\left( \beta \langle \mathcal{E}(\mathbf{k})\rangle + \mathrm{log}\mathcal{Z}(\mathbf{k}) \right),
\end{eqnarray}
where we have used the large gap assumption such that $\langle \mathcal{E}\rangle \approx \mathcal{E}_0$. Using the definition of entropy, we arrive at -
\begin{equation}
    f \approx -\frac{1}{2}S(\mathbf{k}),
\end{equation}
where we have set $k_B = 1$. The Bures metric then becomes
\begin{equation}\bar{g}_{i j}dk^i dk^j  \approx   e^{-S(\mathbf{k})} g_{ i j}dk^i dk^j.  \end{equation}

Note that the above equation is independent of $\beta$ implying that analysis can be extended to more general mixed states, as we show below. 

To this end, we make use of the fact that any density matrix can be expressed as a thermal density matrix in the original eigenbasis but with a thermal Hamiltonian such that,

\begin{equation}
    \rho (\mathbf{k}) = \sum_m \left(e^{-\beta \mathcal{E}_{m, th} (\mathbf{k})}/\mathcal{Z}_{th}\right)|u_m(\mathbf{k})\rangle \langle u_m(\mathbf{k})|,
\end{equation}
where $\mathcal{E}_{m,th}$ are the eigenenergies of the thermal Hamiltonian and not the original Hamiltonian. As before, we assume that $p_0 \gg p_m$ such that Eq. \ref{mixedmetric} becomes,

\begin{equation} 
\bar{g}_{i j}dk^i dk^j \approx \left(e^{-\beta \mathcal{E}_{0,th}}/\mathcal{Z}_{th} \right)  g_{ i j}dk^i dk^j,
\label{mixedmetricthermal}
\end{equation}
and correspondingly,
\begin{equation}
    f_{th} = -\frac{1}{2}\left( \beta \mathcal{E}_{0,th}(\mathbf{k}) + \mathrm{log}\mathcal{Z}_{th}(\mathbf{k}) \right).
\end{equation}
Using the expression for von Neumann entropy $S = - \mathrm{tr}\left( \rho ~\mathrm{log}~\rho \right)$, and the fact that $p_0 \gg p_m$ or $e^{-\beta \mathcal{E}_{0,th}}/\mathcal{Z}_{th} \gg e^{-\beta \mathcal{E}_{m,th}}/\mathcal{Z}_{th}$ $\forall$ $m \neq 0$ implies $\mathcal{E}_{m,th} \gg \mathcal{E}_{m,th}$ or $\langle \mathcal{E}_{th} \rangle \approx \mathcal{E}_{0, th}$, we obtain $f_{th} \approx -\frac{1}{2}S(\rho_{th})$. Writing the von Neumann entropy as $S = -\sum_m p_m ~\mathrm{log}~p_m$, we find that   $S(\rho_{th}) = S(\rho)$ and we again obtain $f \approx -\frac{1}{2}S(\mathbf{k})$. As expected, this result is also independent of $\beta$. In other words, as long as the above-mentioned assumptions are valid, the mixed state can be arbitrary and not just thermal mixed state. 

We see that the mixed state metric differs from the pure state metric by a conformal scale factor due to the von Neumann entropy.  
Because of this simple conformal relation,
the scalar curvature of the Bures metric can be expressed in terms of the scalar curvature of the pure state metric as~\cite{besse2007einstein}:
\begin{equation} \bar{R}=  e^{S/k_B} \left( R +2\Delta_{\mathbf k} S/k_B -\sum_{i}|\partial_{k_i} S/k_B|^2\right), \end{equation}
where $\Delta_{\mathbf k}$ is the Laplace-Beltrami operator associated with the curved momentum space.
Note that we are using the physics convention in which $\Delta$ has a positive sign rather than the math convention that includes an extra factor of $-1$.

Using the trace of the Einstein field equations, we can find the trace of the stress tensor:
\begin{equation} - \frac{1}{2} R  - \frac{1}{2} R' + 3 \Lambda =  T,  \end{equation}
where we have set the typical prefactor of the stress-energy Tensor, $ \frac{8 \pi G}{c^4}$, to one.
The trace of the stress-energy tensor for the Bures metric is thus: 
\begin{equation} T =    - \frac{1}{2}e^{S/k_B} \left( R + 2\Delta_{\mathbf k} S/k_B -\sum_{i}|\partial_{k_i} S/k_B|^2\right) + \frac{1}{2}R. 
\label{stressenergy}
\end{equation}
Assuming S and its first derivatives are small, \eqref{stressenergy} reduces to
\begin{equation}T \approx   -\frac{R}{2 k_B}S(\mathbf k) -   \Delta_{\mathbf k} S(\mathbf k)/k_B . \end{equation}

The full stress-energy tensor can be written as
\begin{equation}
    T_{ij} = -\frac{R}{6 k_{B}} S(\mathbf{k}) g_{ij} + \frac{1}{2 k_{B}} \big[ \nabla_{i} \partial_{j} S(\mathbf{k}) - \mathbf{\Delta} S(\mathbf{k}) g_{ij} \big].
\end{equation}

\subsection{Derivation of Entropic force}

\noindent We know that the equation of motion of a perturbed $n$ band system has contributions from the correction in the ground state energy. For a two band system, the form of the contribution is known. The effect on the equation of motion when the governing metric gets conformally scaled, can be studied by obtaining the Christoffel term using the modified metric. 

\noindent The Christoffel term corresponding to the Bures metric can be expressed in terms of the ones corresponding to the FS metric as,
$$
\Tilde{\Gamma}_{ijl} = e^{2f} \Gamma_{ijl} + \big[ \Tilde{g}_{i j} \partial_{j} f + \Tilde{g}_{il} \partial_{j} f - \Tilde{g}_{jl} \partial_{i} f \big]
$$

\noindent which can be rearranged to obtain,
$$
\Tilde{\Gamma}_{ijl} = e^{2f} \Gamma_{ijl} + \Tilde{g}_{ij} \partial_{l} f
$$
\noindent Now the equation of motion has an additional term, $- \mu \Tilde{g}_{ij} \partial_{l} f E_{0}^{i} E_{0}^{j}$ which can be viewed as an extra force (entropic force) driving the system.

\section{Calculating the Metric and Curvature for a Dirac Fermion}{\label{Appendix D}}

Assuming the charge carrier behaves as a three-dimensional Dirac fermion, we can use the momentum-space Dirac equation as its effective Hamiltonian.
The resulting Bloch function is then described by a four-spinor:
\begin{equation}
|u_0(\mathbf k) \rangle =\textstyle\frac{|\mathcal{E}(\mathbf k)|+\Delta}{\sqrt{\mathbf k^2 + \left(|\mathcal{E}(\mathbf k)| +\Delta \right)^2}} \begin{bmatrix}
    \frac{\vec{\sigma} \cdot \mathbf k}{|\mathcal{E}(\mathbf k)| + \Delta} \chi_s \\ \chi_s 
\end{bmatrix}, 
\end{equation}
where the state has been normalized to one and $|\mathcal{E}(\mathbf k)|$ is the absolute value of the energy of the band, $\mathcal{E}(\mathbf k) = -\sqrt{\Delta^2 + \mathbf{k}^2}$.
Note that $\chi_s$ is an arbitrary two spinor, and the ground state of the Dirac Hamiltonian is thus degenerate.  
While the non-abelian quantum geometric tensor is needed to describe the complete case of degenerate bands, here we focus on the momentum space distance between electron states with the same spin polarization, $\chi_s$.
In this case, the distance in momentum space can be determined from the standard abelian quantum geometric tensor, which itself can be determined from the overlap of the Bloch states via $g_{ij}dk^i dk^j = 1 - |\langle u_0(\mathbf k)|u_0(\mathbf k + \mathbf{dk}) \rangle |^2 $.

Noting the Pauli matrix identity $\left(\vec{a} \cdot \vec{\sigma}\right)\left(\vec{b} \cdot \vec{\sigma}\right) = \left(\vec{a} \cdot \vec{b}\right) \, I + i \left(\vec{a} \times \vec{b}\right) \cdot \vec{\sigma}$ and defining $\mathbf{\hat{s}}$ as the unit vector in the direction of spin polarization, the overlap integral $\langle u_0(\mathbf k)|u_0(\mathbf k') \rangle $ is found to be:

\begin{widetext}
    \begin{equation}
        \begin{gathered}
            \langle u_0(\mathbf k)|u_0(\mathbf k') \rangle =  \textstyle\frac{\left(|\mathcal{E}(\mathbf k)|+\Delta \right)\left(|\mathcal{E'}(\mathbf k')|+\Delta \right)}{\sqrt{\left(\mathbf k^2 + \left(|\mathcal{E}(\mathbf k)| +\Delta \right)^2\right)\left(\mathbf k'^2 + \left(|\mathcal{E'}(\mathbf k')| +\Delta \right)^2\right)}}\left(1 + \frac{\mathbf k \cdot \mathbf k' + i(\mathbf k \times \mathbf k') \cdot\mathbf{\hat{s}}} {\left( \mathcal{E'}(\mathbf k') + \Delta \right)\left( \mathcal{E}(\mathbf k) + \Delta \right)}\right).
        \end{gathered}
    \end{equation}
    
    For the squared magnitude of the overlap we find

\begin{equation}
    \begin{gathered}
        |\langle u_0(\mathbf k)|u_0(\mathbf k') \rangle|^2 =  \textstyle\frac{\left(|\mathcal{E}(\mathbf k)|+\Delta \right)^2\left(|\mathcal{E'}(\mathbf k')|+\Delta \right)^2 + 2\left(|\mathcal{E}(\mathbf k)|+\Delta \right)\left(|\mathcal{E'}(\mathbf k')|+\Delta \right) + \left(\mathbf k \cdot \mathbf k'\right)^2 + \left((\mathbf k \times \mathbf k') \cdot\mathbf{\hat{s}}\right)^2 }{\left(\mathbf k^2 + \left(|\mathcal{E}(\mathbf k)| +\Delta \right)^2\right)\left(\mathbf k'^2 + \left(|\mathcal{E'}(\mathbf k')| +\Delta \right)^2\right)},
    \end{gathered}
\end{equation}

which leads to

\begin{equation}
    \begin{gathered}
        1 - |\langle u_0(\mathbf k)|u_0(\mathbf k') \rangle|^2 =  \textstyle\frac{|\mathbf k \times \mathbf k'|^2-\left((\mathbf k \times \mathbf k') \cdot\mathbf{\hat{s}}\right)^2 + \left|\left(|\mathcal{E'}(\mathbf k')|+\Delta \right) \mathbf k - \left(|\mathcal{E}(\mathbf k)|+\Delta \right)\mathbf k' \right|^2}{\left(\mathbf k^2 + \left(|\mathcal{E}(\mathbf k)| +\Delta \right)^2\right)\left(\mathbf k'^2 + \left(|\mathcal{E'}(\mathbf k')| +\Delta \right)^2\right)}.
    \end{gathered}
\end{equation}

The quantity above can then be re-expressed as:

\begin{equation}
    \begin{gathered}
        1 - |\langle u_0(\mathbf k)|u_0(\mathbf k') \rangle|^2 =  \textstyle\frac{| \frac{\mathbf k}{ |\mathcal{E}(\mathbf k)| +\Delta } \times \frac{\mathbf k'}{ |\mathcal{E'}(\mathbf k')| +\Delta }|^2-\left((\frac{\mathbf k}{ |\mathcal{E}(\mathbf k)| +\Delta } \times \frac{\mathbf k'}{ |\mathcal{E'}(\mathbf k')| +\Delta }) \cdot\mathbf{\hat{s}}\right)^2 + \left|\frac{\mathbf k}{ |\mathcal{E}(\mathbf k)| +\Delta } - \frac{\mathbf k'}{ |\mathcal{E'}(\mathbf k')| +\Delta } \right|^2}{\left(1+ \frac{\mathbf k^2}{\left( |\mathcal{E}(\mathbf k)| +\Delta \right)^2} \right)\left(1+ \frac{\mathbf k'^2}{\left( |\mathcal{E'}(\mathbf k')| +\Delta \right)^2} \right)}.
    \end{gathered}
\end{equation}

\end{widetext}

We assume $\mathbf k^2 \ll \Delta^2$ and expand $1 - |\langle u_0(\mathbf k)|u_0(\mathbf k') \rangle|^2$ in powers of $\frac{\mathbf k}{\Delta }$ and $\frac{\mathbf k'}{ \Delta }$. Keeping only terms of lowest order, we find:
\begin{equation}
1 - |\langle u_0(\mathbf k)|u_0(\mathbf k') \rangle|^2 \approx 
\left|\frac{\mathbf k}{ 2\Delta } - \frac{\mathbf k'}{ 2\Delta } \right|^2 \ \ \ \ \ \ \text{for } \mathbf k^2 \ll \Delta^2. 
\end{equation}
To obtain the metric, we can assume that the separation vector $\mathbf k' - \mathbf k = \mathbf{dk} $, i.e., that the separation of states in momentum space is infinitesimal.
We then find an expression for the quantum geometric tensor when $\mathbf k^2 \ll \Delta$:
\begin{equation}
g_{ij} dk^i dk^j \approx 
 \frac{\delta_{ij} dk^i dk^j }{ 4\Delta ^2}  \ \ \ \ \ \ \text{for } \mathbf k^2 \ll \Delta^2. 
\label{flatmetric}
\end{equation}
We see that the metric is independent of the spin polarization of the fermion states.
Eq.~\eqref{flatmetric} agrees with the expression found previously by Matsuura and Ryu ~\cite{matsuura2010momentum} for the quantum metric of the Dirac fermion.


\begin{thebibliography}{86}
\providecommand{\natexlab}[1]{#1}
\providecommand{\url}[1]{\texttt{#1}}
\expandafter\ifx\csname urlstyle\endcsname\relax
  \providecommand{\doi}[1]{doi: #1}\else
  \providecommand{\doi}{doi: \begingroup \urlstyle{rm}\Url}\fi

\bibitem[Thouless et~al.(1982)Thouless, Kohmoto, Nightingale, and den
  Nijs]{thouless1982quantized}
David~J Thouless, Mahito Kohmoto, M~Peter Nightingale, and Marcel den Nijs.
\newblock Quantized hall conductance in a two-dimensional periodic potential.
\newblock \emph{Physical review letters}, 49\penalty0 (6):\penalty0 405, 1982.

\bibitem[Haldane(1988)]{haldane1988model}
F~Duncan~M Haldane.
\newblock Model for a quantum hall effect without landau levels:
  Condensed-matter realization of the" parity anomaly".
\newblock \emph{Physical review letters}, 61\penalty0 (18):\penalty0 2015,
  1988.

\bibitem[Kane and Mele(2005)]{kane2005quantum}
Charles~L Kane and Eugene~J Mele.
\newblock Quantum spin hall effect in graphene.
\newblock \emph{Physical review letters}, 95\penalty0 (22):\penalty0 226801,
  2005.

\bibitem[Xiao et~al.(2010)Xiao, Chang, and Niu]{xiao2010berry}
Di~Xiao, Ming-Che Chang, and Qian Niu.
\newblock Berry phase effects on electronic properties.
\newblock \emph{Reviews of modern physics}, 82\penalty0 (3):\penalty0 1959,
  2010.

\bibitem[Qi et~al.(2008)Qi, Hughes, and Zhang]{qi2008topological}
Xiao-Liang Qi, Taylor~L Hughes, and Shou-Cheng Zhang.
\newblock Topological field theory of time-reversal invariant insulators.
\newblock \emph{Physical Review B}, 78\penalty0 (19):\penalty0 195424, 2008.

\bibitem[Chang and Niu(1996)]{chang1996berry}
Ming-Che Chang and Qian Niu.
\newblock Berry phase, hyperorbits, and the hofstadter spectrum: Semiclassical
  dynamics in magnetic bloch bands.
\newblock \emph{Physical Review B}, 53\penalty0 (11):\penalty0 7010, 1996.

\bibitem[Sundaram and Niu(1999)]{sundaram1999wave}
Ganesh Sundaram and Qian Niu.
\newblock Wave-packet dynamics in slowly perturbed crystals: Gradient
  corrections and berry-phase effects.
\newblock \emph{Physical Review B}, 59\penalty0 (23):\penalty0 14915, 1999.

\bibitem[Xiao et~al.(2005)Xiao, Shi, and Niu]{xiao2005berry}
Di~Xiao, Junren Shi, and Qian Niu.
\newblock Berry phase correction to electron density of states in solids.
\newblock \emph{Physical review letters}, 95\penalty0 (13):\penalty0 137204,
  2005.

\bibitem[Shi et~al.(2007)Shi, Vignale, Xiao, and Niu]{shi2007quantum}
Junren Shi, Giovanni Vignale, Di~Xiao, and Qian Niu.
\newblock Quantum theory of orbital magnetization and its generalization to
  interacting systems.
\newblock \emph{Physical review letters}, 99\penalty0 (19):\penalty0 197202,
  2007.

\bibitem[Thonhauser et~al.(2005)Thonhauser, Ceresoli, Vanderbilt, and
  Resta]{thonhauser2005orbital}
Timo Thonhauser, Davide Ceresoli, David Vanderbilt, and Raffaele Resta.
\newblock Orbital magnetization in periodic insulators.
\newblock \emph{Physical review letters}, 95\penalty0 (13):\penalty0 137205,
  2005.

\bibitem[Ogata and Fukuyama(2015)]{ogata2015orbital}
Masao Ogata and Hidetoshi Fukuyama.
\newblock Orbital magnetism of bloch electrons i. general formula.
\newblock \emph{Journal of the Physical Society of Japan}, 84\penalty0
  (12):\penalty0 124708, 2015.

\bibitem[Niu(1990)]{niu1990towards}
Q~Niu.
\newblock Towards a quantum pump of electric charges.
\newblock \emph{Physical review letters}, 64\penalty0 (15):\penalty0 1812,
  1990.

\bibitem[Srivastava and Imamo{\u{g}}lu(2015)]{srivastava2015signatures}
Ajit Srivastava and Ata{\c{c}} Imamo{\u{g}}lu.
\newblock Signatures of bloch-band geometry on excitons: nonhydrogenic spectra
  in transition-metal dichalcogenides.
\newblock \emph{Physical review letters}, 115\penalty0 (16):\penalty0 166802,
  2015.

\bibitem[Zhou et~al.(2015)Zhou, Shan, Yao, and Xiao]{zhou2015berry}
Jianhui Zhou, Wen-Yu Shan, Wang Yao, and Di~Xiao.
\newblock Berry phase modification to the energy spectrum of excitons.
\newblock \emph{Physical review letters}, 115\penalty0 (16):\penalty0 166803,
  2015.

\bibitem[Berry(1984)]{berry1984quantal}
Michael~Victor Berry.
\newblock Quantal phase factors accompanying adiabatic changes.
\newblock \emph{Proceedings of the Royal Society of London. A. Mathematical and
  Physical Sciences}, 392\penalty0 (1802):\penalty0 45--57, 1984.

\bibitem[Provost and Vallee(1980)]{provost1980riemannian}
JP~Provost and G~Vallee.
\newblock Riemannian structure on manifolds of quantum states.
\newblock \emph{Communications in Mathematical Physics}, 76\penalty0
  (3):\penalty0 289--301, 1980.

\bibitem[Anandan and Aharonov(1990)]{anandan1990geometry}
J~Anandan and Yakir Aharonov.
\newblock Geometry of quantum evolution.
\newblock \emph{Physical review letters}, 65\penalty0 (14):\penalty0 1697,
  1990.

\bibitem[Marzari et~al.(2012)Marzari, Mostofi, Yates, Souza, and
  Vanderbilt]{marzari2012maximally}
Nicola Marzari, Arash~A Mostofi, Jonathan~R Yates, Ivo Souza, and David
  Vanderbilt.
\newblock Maximally localized wannier functions: Theory and applications.
\newblock \emph{Reviews of Modern Physics}, 84\penalty0 (4):\penalty0 1419,
  2012.

\bibitem[Resta(2011)]{resta2011insulating}
Raffaele Resta.
\newblock The insulating state of matter: a geometrical theory.
\newblock \emph{The European Physical Journal B}, 79\penalty0 (2):\penalty0
  121--137, 2011.

\bibitem[Marzari and Vanderbilt(1997)]{marzari1997maximally}
Nicola Marzari and David Vanderbilt.
\newblock Maximally localized generalized wannier functions for composite
  energy bands.
\newblock \emph{Physical review B}, 56\penalty0 (20):\penalty0 12847, 1997.

\bibitem[Peotta and T{\"o}rm{\"a}(2015)]{peotta2015superfluidity}
Sebastiano Peotta and P{\"a}ivi T{\"o}rm{\"a}.
\newblock Superfluidity in topologically nontrivial flat bands.
\newblock \emph{Nature communications}, 6\penalty0 (1):\penalty0 1--9, 2015.

\bibitem[T{\"o}rm{\"a} et~al.(2018)T{\"o}rm{\"a}, Liang, and
  Peotta]{torma2018quantum}
P{\"a}ivi T{\"o}rm{\"a}, Long Liang, and Sebastiano Peotta.
\newblock Quantum metric and effective mass of a two-body bound state in a flat
  band.
\newblock \emph{Physical Review B}, 98\penalty0 (22):\penalty0 220511, 2018.

\bibitem[Julku et~al.(2021)Julku, Bruun, and
  T{\"o}rm{\"a}]{julku2021excitations}
Aleksi Julku, Georg~M Bruun, and P{\"a}ivi T{\"o}rm{\"a}.
\newblock Excitations of a bose-einstein condensate and the quantum geometry of
  a flat band.
\newblock \emph{arXiv preprint arXiv:2104.14257}, 2021.

\bibitem[Rhim et~al.(2020)Rhim, Kim, and Yang]{rhim2020quantum}
Jun-Won Rhim, Kyoo Kim, and Bohm-Jung Yang.
\newblock Quantum distance and anomalous landau levels of flat bands.
\newblock \emph{Nature}, 584\penalty0 (7819):\penalty0 59--63, 2020.

\bibitem[Chaudhary et~al.(2021)Chaudhary, Lewandowski, and
  Refael]{chaudhary2021shift}
Swati Chaudhary, Cyprian Lewandowski, and Gil Refael.
\newblock Shift-current response as a probe of quantum geometry and
  electron-electron interactions in twisted bilayer graphene.
\newblock \emph{arXiv preprint arXiv:2107.09090}, 2021.

\bibitem[Liang et~al.(2017)Liang, Vanhala, Peotta, Siro, Harju, and
  T{\"o}rm{\"a}]{liang2017band}
Long Liang, Tuomas~I Vanhala, Sebastiano Peotta, Topi Siro, Ari Harju, and
  P{\"a}ivi T{\"o}rm{\"a}.
\newblock Band geometry, berry curvature, and superfluid weight.
\newblock \emph{Physical Review B}, 95\penalty0 (2):\penalty0 024515, 2017.

\bibitem[Lin and Hsiao(2021{\natexlab{a}})]{YuPing2021PRB}
Yu-Ping Lin and Wei-Han Hsiao.
\newblock Dual haldane sphere and quantized band geometry in chiral multifold
  fermions.
\newblock \emph{Phys. Rev. B}, 103:\penalty0 L081103, Feb 2021{\natexlab{a}}.
\newblock \doi{10.1103/PhysRevB.103.L081103}.
\newblock URL \url{https://link.aps.org/doi/10.1103/PhysRevB.103.L081103}.

\bibitem[Rossi(2021)]{rossi2021quantum}
Enrico Rossi.
\newblock Quantum metric and correlated states in two-dimensional systems.
\newblock \emph{arXiv preprint arXiv:2108.11478}, 2021.

\bibitem[Orenstein et~al.(2021)Orenstein, Moore, Morimoto, Torchinsky, Harter,
  and Hsieh]{orenstein2021topology}
J~Orenstein, JE~Moore, T~Morimoto, DH~Torchinsky, JW~Harter, and D~Hsieh.
\newblock Topology and symmetry of quantum materials via nonlinear optical
  responses.
\newblock \emph{Annual Review of Condensed Matter Physics}, 12:\penalty0
  247--272, 2021.

\bibitem[Shi et~al.(2021)Shi, Zhang, Chang, and Song]{shi2021geometric}
Li-kun Shi, Dong Zhang, Kai Chang, and Justin~CW Song.
\newblock Geometric photon-drag effect and nonlinear shift current in
  centrosymmetric crystals.
\newblock \emph{Physical Review Letters}, 126\penalty0 (19):\penalty0 197402,
  2021.

\bibitem[Morimoto and Nagaosa(2016)]{morimoto2016topological}
Takahiro Morimoto and Naoto Nagaosa.
\newblock Topological nature of nonlinear optical effects in solids.
\newblock \emph{Science advances}, 2\penalty0 (5):\penalty0 e1501524, 2016.

\bibitem[Gao et~al.(2014)Gao, Yang, and Niu]{gao2014field}
Yang Gao, Shengyuan~A Yang, and Qian Niu.
\newblock Field induced positional shift of bloch electrons and its dynamical
  implications.
\newblock \emph{Physical review letters}, 112\penalty0 (16):\penalty0 166601,
  2014.

\bibitem[Gao and Xiao(2019)]{gao2019nonreciprocal}
Yang Gao and Di~Xiao.
\newblock Nonreciprocal directional dichroism induced by the quantum metric
  dipole.
\newblock \emph{Physical review letters}, 122\penalty0 (22):\penalty0 227402,
  2019.

\bibitem[Kozii et~al.(2021)Kozii, Avdoshkin, Zhong, and
  Moore]{kozii2021intrinsic}
Vladyslav Kozii, Alexander Avdoshkin, Shudan Zhong, and Joel~E Moore.
\newblock Intrinsic anomalous hall conductivity in a nonuniform electric field.
\newblock \emph{Physical Review Letters}, 126\penalty0 (15):\penalty0 156602,
  2021.

\bibitem[Ahn et~al.(2021)Ahn, Guo, Nagaosa, and Vishwanath]{ahn2021riemannian}
Junyeong Ahn, Guang-Yu Guo, Naoto Nagaosa, and Ashvin Vishwanath.
\newblock Riemannian geometry of resonant optical responses.
\newblock \emph{arXiv preprint arXiv: 2103.01241}, 2021.

\bibitem[Sodemann and Fu(2015)]{sodemann2015quantum}
Inti Sodemann and Liang Fu.
\newblock Quantum nonlinear hall effect induced by berry curvature dipole in
  time-reversal invariant materials.
\newblock \emph{Physical review letters}, 115\penalty0 (21):\penalty0 216806,
  2015.

\bibitem[Topp et~al.(2021)Topp, Eckhardt, Kennes, Sentef, and
  T{\"o}rm{\"a}]{topp2021light}
Gabriel~E Topp, Christian~J Eckhardt, Dante~M Kennes, Michael~A Sentef, and
  P{\"a}ivi T{\"o}rm{\"a}.
\newblock Light-matter coupling and quantum geometry in moir$\backslash$'e
  materials.
\newblock \emph{arXiv preprint arXiv:2103.04967}, 2021.

\bibitem[Claassen et~al.(2015)Claassen, Lee, Thomale, Qi, and
  Devereaux]{claassen2015position}
Martin Claassen, Ching~Hua Lee, Ronny Thomale, Xiao-Liang Qi, and Thomas~P
  Devereaux.
\newblock Position-momentum duality and fractional quantum hall effect in chern
  insulators.
\newblock \emph{Physical review letters}, 114\penalty0 (23):\penalty0 236802,
  2015.

\bibitem[Haldane(2011)]{haldane2011geometrical}
FDM Haldane.
\newblock Geometrical description of the fractional quantum hall effect.
\newblock \emph{Physical review letters}, 107\penalty0 (11):\penalty0 116801,
  2011.

\bibitem[Palumbo(2018)]{Palumbo2018EPL}
Giandomenico Palumbo.
\newblock Momentum-space cigar geometry in topological phases.
\newblock \emph{The European Physical Journal Plus}, 133\penalty0 (1):\penalty0
  23, Jan 2018.
\newblock ISSN 2190-5444.
\newblock \doi{10.1140/epjp/i2018-11856-8}.
\newblock URL \url{https://doi.org/10.1140/epjp/i2018-11856-8}.

\bibitem[Salerno et~al.(2020)Salerno, Goldman, and Palumbo]{Palumbo2020PRX}
Grazia Salerno, Nathan Goldman, and Giandomenico Palumbo.
\newblock Floquet-engineering of nodal rings and nodal spheres and their
  characterization using the quantum metric.
\newblock \emph{Phys. Rev. Research}, 2:\penalty0 013224, Feb 2020.
\newblock \doi{10.1103/PhysRevResearch.2.013224}.
\newblock URL \url{https://link.aps.org/doi/10.1103/PhysRevResearch.2.013224}.

\bibitem[Lin and Hsiao(2021{\natexlab{b}})]{lin2021band}
Yu-Ping Lin and Wei-Han Hsiao.
\newblock Band geometry from position-momentum duality at topological band
  crossings.
\newblock \emph{arXiv preprint arXiv:2102.04470}, 2021{\natexlab{b}}.

\bibitem[Neupert et~al.(2013)Neupert, Chamon, and Mudry]{neupert2013measuring}
Titus Neupert, Claudio Chamon, and Christopher Mudry.
\newblock Measuring the quantum geometry of bloch bands with current noise.
\newblock \emph{Physical Review B}, 87\penalty0 (24):\penalty0 245103, 2013.

\bibitem[Pi{\'e}chon et~al.(2016)Pi{\'e}chon, Raoux, Fuchs, and
  Montambaux]{piechon2016geometric}
Fr{\'e}d{\'e}ric Pi{\'e}chon, Arnaud Raoux, Jean-No{\"e}l Fuchs, and Gilles
  Montambaux.
\newblock Geometric orbital susceptibility: Quantum metric without berry
  curvature.
\newblock \emph{Physical Review B}, 94\penalty0 (13):\penalty0 134423, 2016.

\bibitem[Zanardi et~al.(2008)Zanardi, Paris, and Venuti]{zanardi2008quantum}
Paolo Zanardi, Matteo~GA Paris, and Lorenzo~Campos Venuti.
\newblock Quantum criticality as a resource for quantum estimation.
\newblock \emph{Physical Review A}, 78\penalty0 (4):\penalty0 042105, 2008.

\bibitem[Ma et~al.(2010)Ma, Chen, Fan, Liu, et~al.]{ma2010abelian}
Yu-Quan Ma, Shu Chen, Heng Fan, Wu-Ming Liu, et~al.
\newblock Abelian and non-abelian quantum geometric tensor.
\newblock \emph{Physical Review B}, 81\penalty0 (24):\penalty0 245129, 2010.

\bibitem[Rezakhani et~al.(2010)Rezakhani, Abasto, Lidar, and
  Zanardi]{rezakhani2010intrinsic}
Ali~T Rezakhani, Damian~F Abasto, Daniel~A Lidar, and Paolo Zanardi.
\newblock Intrinsic geometry of quantum adiabatic evolution and quantum phase
  transitions.
\newblock \emph{Physical Review A}, 82\penalty0 (1):\penalty0 012321, 2010.

\bibitem[Zanardi et~al.(2007)Zanardi, Giorda, and
  Cozzini]{zanardi2007information}
Paolo Zanardi, Paolo Giorda, and Marco Cozzini.
\newblock Information-theoretic differential geometry of quantum phase
  transitions.
\newblock \emph{Physical review letters}, 99\penalty0 (10):\penalty0 100603,
  2007.

\bibitem[Gianfrate et~al.(2020)Gianfrate, Bleu, Dominici, Ardizzone, De~Giorgi,
  Ballarini, Lerario, West, Pfeiffer, Solnyshkov,
  et~al.]{gianfrate2020measurement}
A~Gianfrate, O~Bleu, L~Dominici, V~Ardizzone, M~De~Giorgi, D~Ballarini,
  G~Lerario, KW~West, LN~Pfeiffer, DD~Solnyshkov, et~al.
\newblock Measurement of the quantum geometric tensor and of the anomalous hall
  drift.
\newblock \emph{Nature}, 578\penalty0 (7795):\penalty0 381--385, 2020.

\bibitem[Ozawa and Goldman(2018)]{ozawa2018extracting}
Tomoki Ozawa and Nathan Goldman.
\newblock Extracting the quantum metric tensor through periodic driving.
\newblock \emph{Physical Review B}, 97\penalty0 (20):\penalty0 201117, 2018.

\bibitem[Price et~al.(2014)Price, Ozawa, and Carusotto]{price2014quantum}
Hannah~M Price, Tomoki Ozawa, and Iacopo Carusotto.
\newblock Quantum mechanics with a momentum-space artificial magnetic field.
\newblock \emph{Physical review letters}, 113\penalty0 (19):\penalty0 190403,
  2014.

\bibitem[Price et~al.(2015)Price, Zilberberg, Ozawa, Carusotto, and
  Goldman]{price2015four}
Hannah~M Price, Oded Zilberberg, Tomoki Ozawa, Iacopo Carusotto, and Nathan
  Goldman.
\newblock Four-dimensional quantum hall effect with ultracold atoms.
\newblock \emph{Physical review letters}, 115\penalty0 (19):\penalty0 195303,
  2015.

\bibitem[Fang et~al.(2003)Fang, Nagaosa, Takahashi, Asamitsu, Mathieu,
  Ogasawara, Yamada, Kawasaki, Tokura, and Terakura]{fang2003anomalous}
Zhong Fang, Naoto Nagaosa, Kei~S Takahashi, Atsushi Asamitsu, Roland Mathieu,
  Takeshi Ogasawara, Hiroyuki Yamada, Masashi Kawasaki, Yoshinori Tokura, and
  Kiyoyuki Terakura.
\newblock The anomalous hall effect and magnetic monopoles in momentum space.
\newblock \emph{Science}, 302\penalty0 (5642):\penalty0 92--95, 2003.

\bibitem[Ozawa et~al.(2015)Ozawa, Price, and Carusotto]{ozawa2015momentum}
Tomoki Ozawa, Hannah~M Price, and Iacopo Carusotto.
\newblock Momentum-space harper-hofstadter model.
\newblock \emph{Physical Review A}, 92\penalty0 (2):\penalty0 023609, 2015.

\bibitem[Yang et~al.(2012)Yang, Hu, Papi\ifmmode~\acute{c}\else \'{c}\fi{}, and
  Haldane]{HaldaneGraviton2012PRL}
Bo~Yang, Zi-Xiang Hu, Z.~Papi\ifmmode~\acute{c}\else \'{c}\fi{}, and F.~D.~M.
  Haldane.
\newblock Model wave functions for the collective modes and the magnetoroton
  theory of the fractional quantum hall effect.
\newblock \emph{Phys. Rev. Lett.}, 108:\penalty0 256807, Jun 2012.
\newblock \doi{10.1103/PhysRevLett.108.256807}.
\newblock URL \url{https://link.aps.org/doi/10.1103/PhysRevLett.108.256807}.

\bibitem[Golkar et~al.(2016)Golkar, Nguyen, and Son]{Golkar2016JHP}
Siavash Golkar, Dung~X. Nguyen, and Dam~T. Son.
\newblock Spectral sum rules and magneto-roton as emergent graviton in
  fractional quantum hall effect.
\newblock \emph{Journal of High Energy Physics}, 2016\penalty0 (1):\penalty0
  21, Jan 2016.
\newblock ISSN 1029-8479.
\newblock \doi{10.1007/JHEP01(2016)021}.
\newblock URL \url{https://doi.org/10.1007/JHEP01(2016)021}.

\bibitem[Davis and Foster(2021)]{davis2021geodesic}
Seth~M. Davis and Matthew~S. Foster.
\newblock Geodesic geometry of 2+1-d dirac materials subject to artificial,
  quenched gravitational singularities.
\newblock \emph{arXiv preprint arXiv:2107.04047}, 2021.

\bibitem[Kirmani et~al.(2021)Kirmani, Bull, Hou, Papić, Rahmani, and
  Ghaemi]{kirmani2021realizing}
Ammar Kirmani, Kieran Bull, Chang-Yu Hou, Zlatko Papić, Armin Rahmani, and
  Pouyan Ghaemi.
\newblock Realizing fractional-quantum-hall gravitons on quantum computers.
\newblock \emph{arXiv preprint arXiv:2107.10267}, 2021.

\bibitem[Wilson et~al.(2020)Wilson, Curtis, and Galitski]{wilson2020analogue}
Justin~H. Wilson, Jonathan~B. Curtis, and Victor~M. Galitski.
\newblock Analogue spacetimes from nonrelativistic goldstone modes in spinor
  condensates.
\newblock \emph{arXiv preprint arXiv:2001.05496}, 2020.

\bibitem[Carroll(2019)]{carroll2019spacetime}
Sean~M Carroll.
\newblock \emph{Spacetime and geometry}.
\newblock Cambridge University Press, 2019.

\bibitem[Cao et~al.(2018)Cao, Fatemi, Demir, Fang, Tomarken, Luo,
  Sanchez-Yamagishi, Watanabe, Taniguchi, Kaxiras, et~al.]{cao2018correlated}
Yuan Cao, Valla Fatemi, Ahmet Demir, Shiang Fang, Spencer~L Tomarken, Jason~Y
  Luo, Javier~D Sanchez-Yamagishi, Kenji Watanabe, Takashi Taniguchi, Efthimios
  Kaxiras, et~al.
\newblock Correlated insulator behaviour at half-filling in magic-angle
  graphene superlattices.
\newblock \emph{Nature}, 556\penalty0 (7699):\penalty0 80--84, 2018.

\bibitem[Naik and Jain(2018)]{naik2018ultraflatbands}
Mit~H Naik and Manish Jain.
\newblock Ultraflatbands and shear solitons in moir{\'e} patterns of twisted
  bilayer transition metal dichalcogenides.
\newblock \emph{Physical review letters}, 121\penalty0 (26):\penalty0 266401,
  2018.

\bibitem[Bekenstein(1973)]{bekenstein1973black}
Jacob~D Bekenstein.
\newblock Black holes and entropy.
\newblock \emph{Physical Review D}, 7\penalty0 (8):\penalty0 2333, 1973.

\bibitem[Hawking(1975)]{hawking1975particle}
Stephen~W Hawking.
\newblock Particle creation by black holes.
\newblock \emph{Communications in mathematical physics}, 43\penalty0
  (3):\penalty0 199--220, 1975.

\bibitem[Ruppeiner(1979)]{ruppeiner1979thermodynamics}
George Ruppeiner.
\newblock Thermodynamics: A riemannian geometric model.
\newblock \emph{Physical Review A}, 20\penalty0 (4):\penalty0 1608, 1979.

\bibitem[Jacobson(1995)]{jacobson1995thermodynamics}
Ted Jacobson.
\newblock Thermodynamics of spacetime: the einstein equation of state.
\newblock \emph{Physical Review Letters}, 75\penalty0 (7):\penalty0 1260, 1995.

\bibitem[Padmanabhan(2010)]{padmanabhan2010thermodynamical}
Thanu Padmanabhan.
\newblock Thermodynamical aspects of gravity: new insights.
\newblock \emph{Reports on Progress in Physics}, 73\penalty0 (4):\penalty0
  046901, 2010.

\bibitem[Verlinde(2011)]{verlinde2011origin}
Erik Verlinde.
\newblock On the origin of gravity and the laws of newton.
\newblock \emph{Journal of High Energy Physics}, 2011\penalty0 (4):\penalty0
  1--27, 2011.

\bibitem[Maldacena and Susskind(2013)]{maldacena2013cool}
Juan Maldacena and Leonard Susskind.
\newblock Cool horizons for entangled black holes.
\newblock \emph{Fortschritte der Physik}, 61\penalty0 (9):\penalty0 781--811,
  2013.

\bibitem[Van~Raamsdonk(2010)]{van2010building}
Mark Van~Raamsdonk.
\newblock Building up spacetime with quantum entanglement.
\newblock \emph{General Relativity and Gravitation}, 42\penalty0 (10):\penalty0
  2323--2329, 2010.

\bibitem[Misner et~al.(1973)Misner, Thorne, Wheeler,
  et~al.]{misner1973gravitation}
Charles~W Misner, Kip~S Thorne, John~Archibald Wheeler, et~al.
\newblock \emph{Gravitation}.
\newblock Macmillan, 1973.

\bibitem[Chang and Niu(2008)]{chang2008berry}
Ming-Che Chang and Qian Niu.
\newblock Berry curvature, orbital moment, and effective quantum theory of
  electrons in electromagnetic fields.
\newblock \emph{Journal of Physics: Condensed Matter}, 20\penalty0
  (19):\penalty0 193202, 2008.

\bibitem[Xiao et~al.(2021{\natexlab{a}})Xiao, Liu, Zhao, Yang, and
  Niu]{Xiao2021PRBthermo}
Cong Xiao, Huiying Liu, Jianzhou Zhao, Shengyuan~A. Yang, and Qian Niu.
\newblock Thermoelectric generation of orbital magnetization in metals.
\newblock \emph{Phys. Rev. B}, 103:\penalty0 045401, Jan 2021{\natexlab{a}}.
\newblock \doi{10.1103/PhysRevB.103.045401}.
\newblock URL \url{https://link.aps.org/doi/10.1103/PhysRevB.103.045401}.

\bibitem[Xiao et~al.(2021{\natexlab{b}})Xiao, Ren, and Xiong]{Xiao2021PRBmag}
Cong Xiao, Yafei Ren, and Bangguo Xiong.
\newblock Adiabatically induced orbital magnetization.
\newblock \emph{Phys. Rev. B}, 103:\penalty0 115432, Mar 2021{\natexlab{b}}.
\newblock \doi{10.1103/PhysRevB.103.115432}.
\newblock URL \url{https://link.aps.org/doi/10.1103/PhysRevB.103.115432}.

\bibitem[Gao et~al.(2015)Gao, Yang, and Niu]{gao2015geometrical}
Yang Gao, Shengyuan~A Yang, and Qian Niu.
\newblock Geometrical effects in orbital magnetic susceptibility.
\newblock \emph{Physical Review B}, 91\penalty0 (21):\penalty0 214405, 2015.

\bibitem[Koshino et~al.(2018)Koshino, Yuan, Koretsune, Ochi, Kuroki, and
  Fu]{Koshino2018PRX}
Mikito Koshino, Noah F.~Q. Yuan, Takashi Koretsune, Masayuki Ochi, Kazuhiko
  Kuroki, and Liang Fu.
\newblock Maximally localized wannier orbitals and the extended hubbard model
  for twisted bilayer graphene.
\newblock \emph{Phys. Rev. X}, 8:\penalty0 031087, Sep 2018.
\newblock \doi{10.1103/PhysRevX.8.031087}.
\newblock URL \url{https://link.aps.org/doi/10.1103/PhysRevX.8.031087}.

\bibitem[Moroianu(2007)]{moroianu2007lectures}
Andrei Moroianu.
\newblock \emph{Lectures on K{\"a}hler geometry}, volume~69.
\newblock Cambridge University Press, 2007.

\bibitem[Besse(2007)]{besse2007einstein}
Arthur~L Besse.
\newblock \emph{Einstein manifolds}.
\newblock Springer Science \& Business Media, 2007.

\bibitem[foo()]{footnote}
While the momentum space inherits the Hilbert-space metric, the curvature of
  this subspace may be different due to the reduction in dimension. A similar
  situation arises in general relativity: the intrinsic curvature on a
  hypersurface may be different than the curvature of the full
  spacetime~\cite{baez1994gauge}. The momentum-space stress-energy can thus be
  nonzero due to curvature associated with the embedding of momentum space in a
  higher-dimensional Hilbert space.

\bibitem[Bures(1969)]{bures1969extension}
Donald Bures.
\newblock An extension of kakutani's theorem on infinite product measures to
  the tensor product of semifinite w*-algebras.
\newblock \emph{Transactions of the American Mathematical Society},
  135:\penalty0 199--212, 1969.

\bibitem[Dittmann(1993)]{dittmann1993riemannian}
Jochen Dittmann.
\newblock On the riemannian geometry of finite dimensional mixed states.
\newblock In \emph{Seminar Sophus Lie}, volume~3, page~2. Citeseer, 1993.

\bibitem[Matsuura and Ryu(2010)]{matsuura2010momentum}
Shunji Matsuura and Shinsei Ryu.
\newblock Momentum space metric, nonlocal operator, and topological insulators.
\newblock \emph{Physical Review B}, 82\penalty0 (24):\penalty0 245113, 2010.

\bibitem[Ikeda(2020)]{ikeda2020high}
Tatsuhiko~N Ikeda.
\newblock High-order nonlinear optical response of a twisted bilayer graphene.
\newblock \emph{Physical Review Research}, 2\penalty0 (3):\penalty0 032015,
  2020.

\bibitem[Wu and Sarma(2020)]{wu2020quantum}
Fengcheng Wu and S~Das Sarma.
\newblock Quantum geometry and stability of moir{\'e} flatband ferromagnetism.
\newblock \emph{Physical Review B}, 102\penalty0 (16):\penalty0 165118, 2020.

\bibitem[Dittmann(1999)]{dittmann1999explicit}
J~Dittmann.
\newblock Explicit formulae for the bures metric.
\newblock \emph{Journal of Physics A: Mathematical and General}, 32\penalty0
  (14):\penalty0 2663, 1999.

\bibitem[Baez and Muniain(1994)]{baez1994gauge}
John~C Baez and Javier~P Muniain.
\newblock \emph{Gauge fields, knots and gravity}, volume~4.
\newblock World Scientific Publishing Company, 1994.

\end{thebibliography}

\end{document}